\documentclass[namedreferences]{solarphysics}

\usepackage[hyperref,optionalrh]{spr-sola-addons} 
\usepackage{graphicx}        
\usepackage{color}           
\usepackage[normalem]{ulem}

\chardef\us=`\_

\begin{document}
\begin{article}

\begin{opening}

\title{Three-Dimensional Reconstructions of Coronal Wave Surfaces Using a New Mask-Fitting Method}

\author[addressref=aff1,corref,email={lfeng@pmo.ac.cn}]{\inits{L.}\fnm{Li}~\lnm{Feng}
\orcid{0003-4655-6939}}
\author[addressref=aff1,email={leilu@pmo.ac.cn}]{\inits{L.}\fnm{Lei}~\lnm{Lu}
\orcid{0002-3032-6066}}
\author[addressref=aff2,email={binhest@mps.mpg.de}]{\inits{B.}\fnm{Bernd}~\lnm{Inhester}}
\author[addressref=aff3,email={jplowman@nso.edu}]{\inits{J.}\fnm{Joseph}~\lnm{Plowman}
\orcid{0001-7016-7226}}
\author[addressref=aff1,email={yingbl@pmo.ac.cn}]{\inits{B.}\fnm{Beili}~\lnm{Ying}
\orcid{0001-8402-9748}}
\author[addressref=aff4,email={marilena.mierla@oma.be}]{\inits{M.}\fnm{Marilena}~\lnm{Mierla}
\orcid{0003-4105-7364}}
\author[addressref=aff4,email={mwest@oma.be}]{\inits{M.J.}\fnm{Matthew J.}~\lnm{West}
\orcid{0002-0631-2393}}
\author[addressref=aff1,email={wqgan@pmo.ac.cn}]{\inits{W.}\fnm{Weiqun}~\lnm{Gan}}


\address[id=aff1]{Key Laboratory of Dark Matter and Space Astronomy, Purple Mountain Observatory,
Chinese Academy of Sciences, Nanjing 210033, Jiangsu, China}
\address[id=aff2]{Max-Planck-Institut f\"{u}r Sonnensystemforschung, G\"{o}ttingen 37077, Lower Saxony, Germany}
\address[id=aff3]{National Solar Observatory, Boulder, CO 80303, USA}
\address[id=aff4]{Royal Observatory of Belgium, 1180 Brussels, Belgium}

\runningauthor{L. Feng et al.}
\runningtitle{3D Coronal Wave Surfaces}

\begin{abstract}
Coronal waves are large-scale disturbances often driven by coronal mass ejections (CMEs). We investigate a spectacular wave event on 7 March 2012, which is associated with an X5.4 flare (SOL2012-03-07). By using a running center-median (RCM) filtering method for the detection of temporal variations in extreme ultraviolet (EUV) images, we enhance the EUV disturbance observed by the \textit{Atmospheric Imaging Assembly} (AIA) onboard the \textit{Solar Dynamics Observatory} (SDO) and the \textit{Sun Watcher using Active Pixel System detector and Image Processing} (SWAP) onboard the \textit{PRoject for Onboard Autonomy 2} (PROBA2). In coronagraph images, a halo front is observed to be the upper counterpart of the EUV disturbance. Based on the EUV and coronagraph images observed from three different perspectives, we have made three-dimensional (3D) reconstructions of the wave surfaces using a new mask-fitting method. The reconstructions are compared with those obtained from forward-fitting methods. We show that the mask fitting method can reflect the inhomogeneous coronal medium by capturing the concave shape of the shock wave front. Subsequently, we trace the developing concave structure and derive the deprojected wave kinematics. The speed of the 3D-wave nose increases from a low value below a few hundred km~s$^{-1}$ to a maximum value of about 3800\,km~s$^{-1}$, and then slowly decreases afterwards.  The concave structure starts to decelerate earlier and has significantly lower speeds than those of the wave nose. We also find that the 3D-wave in the extended corona has a much higher speed than the speed of EUV disturbances across the solar disk.

\end{abstract}
\keywords{Waves, Corona; Coronal mass ejections, Corona; Image Processing, Techniques}
\end{opening}

\section{Introduction}

Large-scale coronal propagating fronts in extreme ultraviolet (EUV)  are often driven by the disturbances of coronal mass ejections (CMEs) or flares. They were first analyzed using the \textit{Extreme ultraviolet Imaging Telescope} (EIT) images in the late 1990s \citep{Moses1997, Thompson1998}. After the launch of the \textit{Solar Dynamics Observatory} (SDO) in 2010,  
EUV images observed by the \textit{Atmospheric Imaging Assembly} (AIA) provide an improved view of coronal waves \citep[e.g.][]{ShenYD2012, Nitta2013}. The authors measured the speed of propagating fronts in various directions in AIA images, and they found that their highest speed was often considerably higher than the speed measured in EIT images. Moreover, the obtained speeds were not closely correlated with the flare class and CME speed, nor were they strongly associated with Type II radio bursts. 

Coronal propagating fronts have now been observed in wide spectral ranges not only in EUV, but also in soft X-ray, white light,  and radio (see \citealp{Warmuth2015}, for a review).  They sometimes have a chromospheric imprint \citep{Uchida1968} observed in H$\alpha$ as Moreton waves \citep[e.g.][]{Moreton1960} and in He\,{\scshape i} at 10830 \AA\,\citep[e.g.][]{Vrsnak2002}. Concerning the physical interpretation of these disturbances, many observations suggest that they are fast-mode magnetohydrodynamic (MHD) waves. However, some other models have also been proposed, e.g. the hybrid model proposed by \citet{ChenPF2002} with the first front being a wave, and the second front formed due to magnetic reconfiguration. For a more complete discussion of the models, please refer to \citet{Warmuth2015}. Due to their different physical interpretations,  in this article, we call the propagating fronts in EUV ``EUV disturbances". They can be a true wave (a linear fast-mode MHD wave or a nonlinear shock wave) or a wave-like front due to magnetic reconfiguration.


The coronal disturbances in the EUV and white-light coronagraph images have diverse morphologies. On-disk, bright, propagating fronts are followed by dimming regions behind \citep{Thompson1998}. Off-limb, in uncomplicated magnetic regions, large-scale coronal EUV waves can be seen to expand as dome-shaped bubbles, first described by \citet{Veronig2010}. The structures beyond the limb often perfectly connect to the on-disk structures.  
It is also found by \citet{Morosan2019} that the EUV wave can be regarded as the flank of the wave bubble that is extending upwards in the high corona.
In white-light coronagraph images, shocks are manifested as a second front preceding the CME front \citep[e.g.][]{Vourlidas2003, Ontiveros2009, Feng2012}.
\citet{Kwon2014} devised a three-dimensional (3D) ellipsoidal model to fit the shock wave observed from multi-perspectives of \textit{Solar TErrestrial RElations Observatory} (STEREO)-A (STA),  STEREO-B (STB), and Earth from EUV to coronagraph field of view (FOV).
Understanding 3D-structures of the wave is crucial to unlock their physical nature. A precise localization of the 3D-shock surface is important to characterize the shock parameters, e.g. shock velocity, the angle between shock normal and magnetic field. They shed important light on the particle acceleration by shocks, and they reveal the relationship between shock parameters and solar energetic particles (SEPs) \citep{Lario2014, Kozarev2015, Lario2016, Lario2017a, Lario2017b}.

In this article, we investigate the eruption event on 7 March 2012, which was associated with an X5.4 flare (SOL2012-03-07). Due to its spectacular EUV disturbances almost across the entire solar disk, very high CME speed, and large flare class, the event has drawn much attention. Researches have been carried out on the properties of magnetic flux rope \citep[e.g.][]{ZhouGP2019}, EUV disturbances \citep[e.g.][]{Takahashi2015}, CME, and shock propagation \citep[e.g.][]{Kwon2014, Rollett2014, ZhaoXH2016, Patsourakos2016}, SEP characteristics \citep[e.g.][]{Kouloumvakos2016, DingLG2016}, long-duration gamma-ray emissions \citep[e.g.][]{ Ajello2014}, etc.

We take advantage of the large FOV of the \textit{Sun Watcher using Active pixel system detector and image Processing} (SWAP: \citealp{Seaton2013}; \citealp{Halain2013}), which extends to about 1.7~$\mathrm{R_\odot}$, to have a more complete view of the EUV disturbances.  The observations of the disturbances in EUV and white-light coronagraph images, and possibly associated Type II emission are presented in Section~2. In Section~3, we describe the running center-median filtering technique to improve the visibility of the EUV disturbances. The 3D-reconstructions of the coronal wave surface using the improved mask-fitting method is also introduced and evaluated. Section~4 includes the results of 3D-reconstructions using mask-fitting and forward fitting methods and their comparisons. Further analyses of wave morphology and kinematics are also presented. The 3D-kinematics in the extended corona is subsequently compared with the kinematics on the solar disk as seen by AIA and with that derived from the associated Type II burst. In the last section are conclusions and outlook.

\section{Observations}


\subsection{EUV and Coronagraph Observations from Three Perspectives}

\begin{figure}[htbp]
\begin{center}
\vbox{
\includegraphics[width=1.\textwidth,clip=]{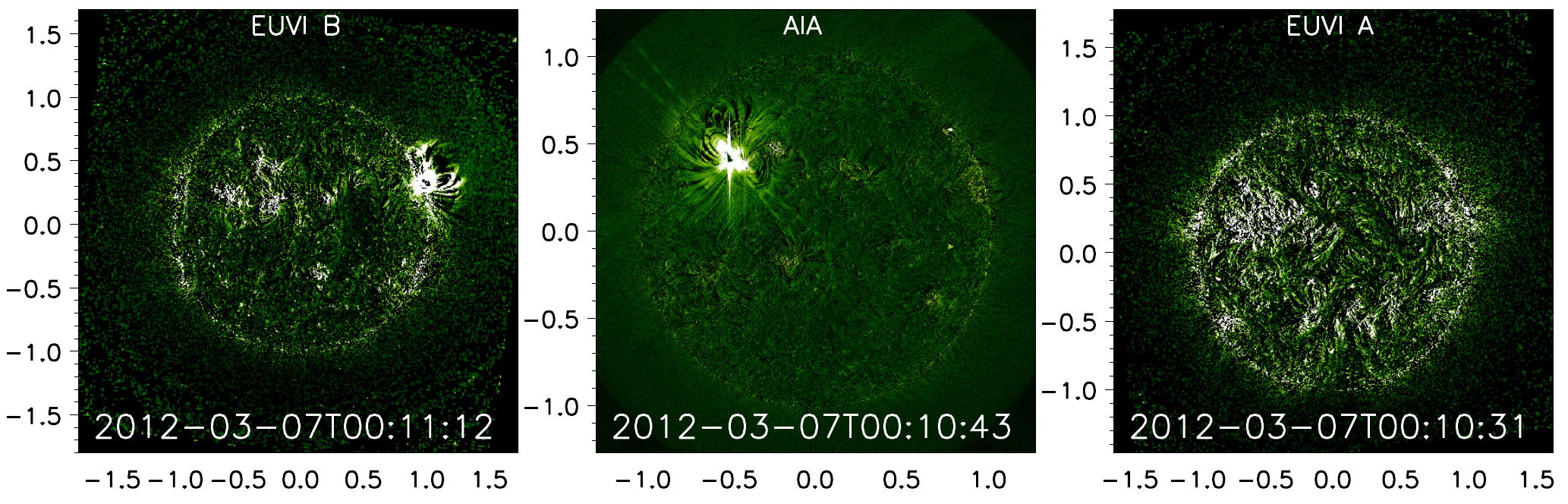}
\includegraphics[width=1.\textwidth,clip=]{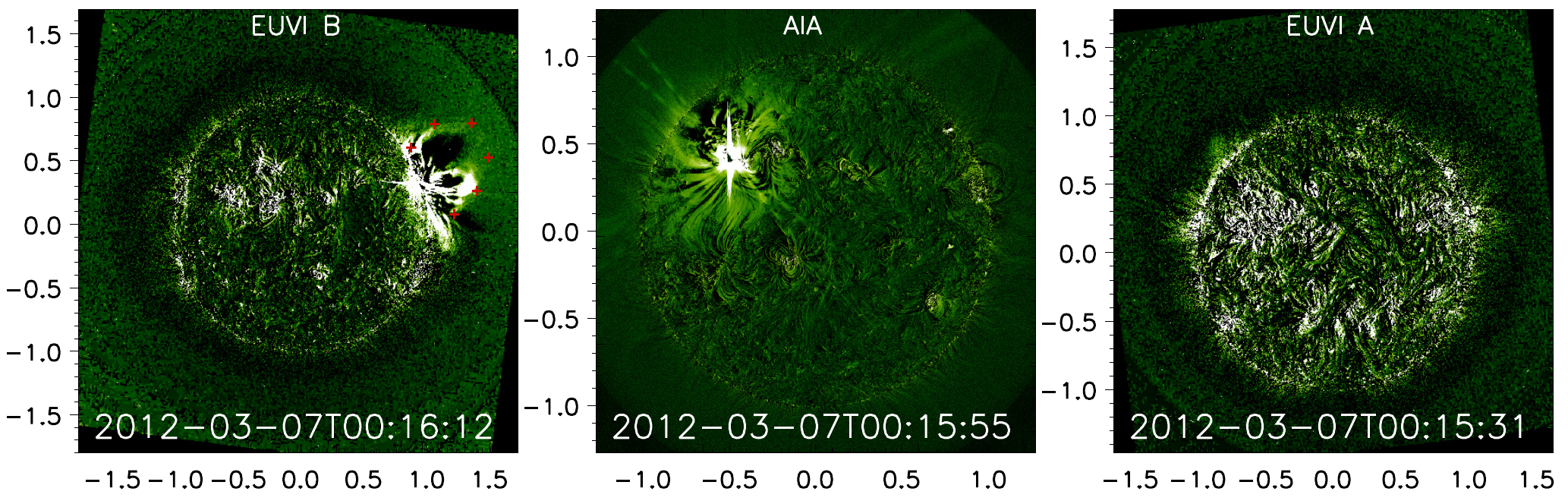}
\includegraphics[width=1.\textwidth,clip=]{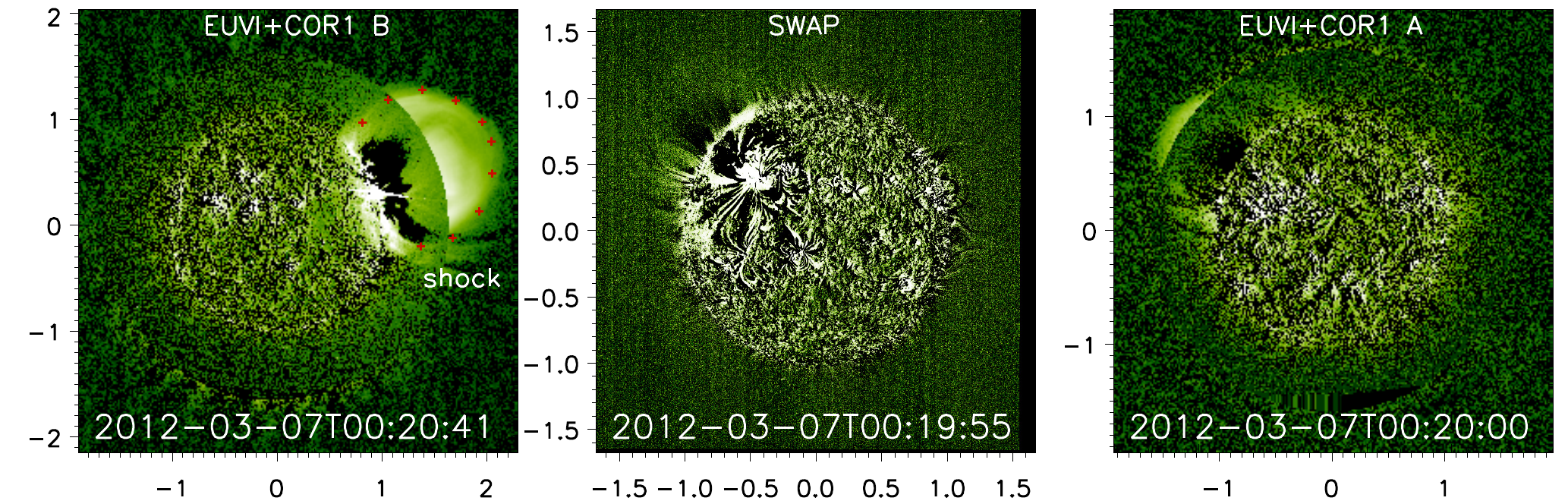}
\includegraphics[width=1\textwidth,clip=]{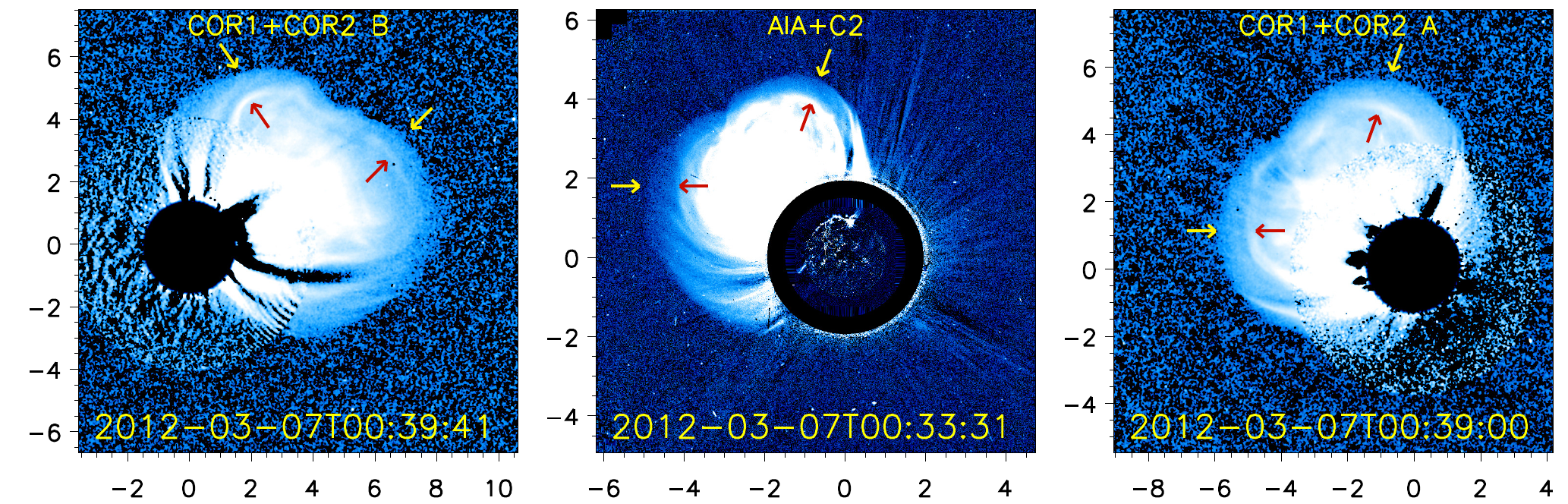}
\includegraphics[width=1\textwidth,clip=]{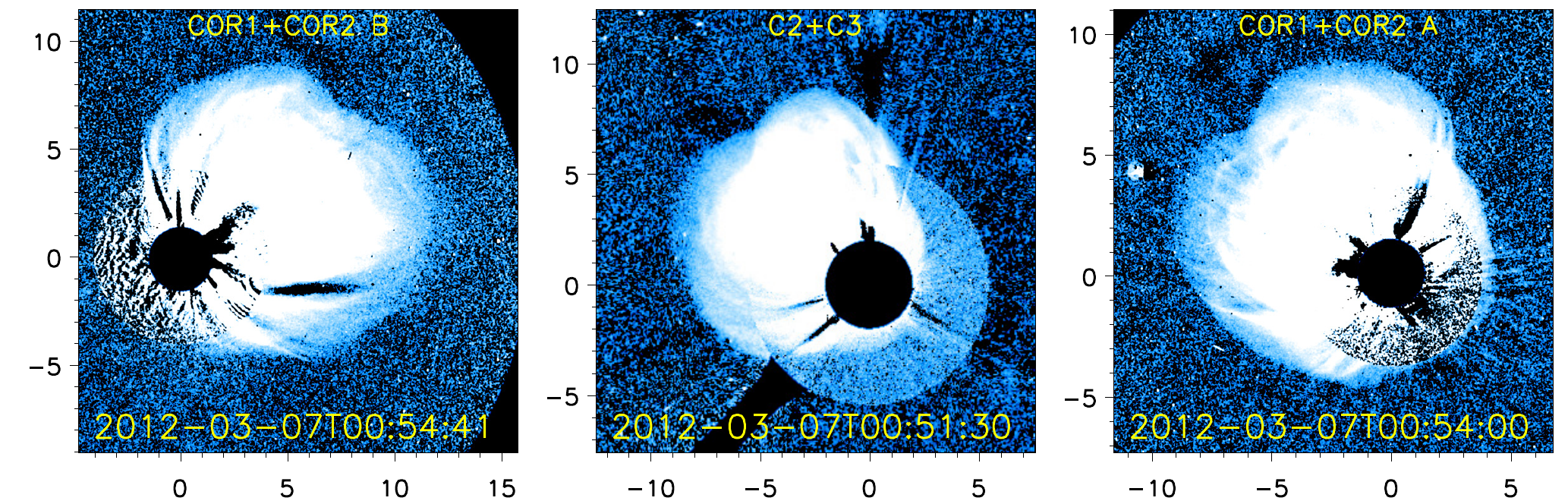}
}
\caption{The observations in EUV and white-light coronagraph images at five representative times taken by STB and STA are presented in the left and right columns, respectively. The corresponding observations by SDO/AIA, PROBA 2/SWAP, SOHO/LASCO  at a closer time from the third perspective are displayed in the middle column. All images are base-differerence images. The red-plus (+) symbols in the left panel indicate the CME front observed by STB. The red arrows in the fourth row point to the CME front, while the yellow arrows point to the shock front driven by the CME. All of the axes are in units of solar radius. }
\label{fig:euv_cor_3view}
\end{center}
\end{figure}

The event that we are investigating occurred on 7 March 2012 and was associated with an X5.4 flare.  A large-scale EUV disturbance was observed and propagated almost across the entire solar disk from the perspective of the Earth (see an Electronic Supplementary Material of the wave propagation observed in AIA at 193\,\AA). The event was seen in AIA, with a FOV of about 1.2\,$\mathrm{R_\odot}$ and a temporal cadence of 12~seconds. This event was also observed by STA and STB from different perspectives. The positions of STA, Earth, and STB were separated by about 120 degrees from each other. The \textit{Extreme UltraViolet Imager} (EUVI) of the \textit{Sun Earth Connection Coronal and Heliospheric Investigation} (SECCHI) instrument suite onboard STA and STB have a FOV of about 1.6\,$\mathrm{R_\odot}$. However, the time cadence of EUVI is much lower, and is about five minutes at 195\,\AA\,for this event. The triple-view observations at about 00:11~UT and 00:16~UT are presented in the first and second rows of Figure~\ref{fig:euv_cor_3view}. For observations of EUV disturbances, a preferable emission temperature is about $1.5\times10^6$~K.  Therefore, observations from EUVI at 195\,\AA\, and from AIA at 193\,\AA\, are shown in Figure~\ref{fig:euv_cor_3view}, as the corresponding passbands peak near this temperature. All images are base-difference images. Note that at about 00:16~UT as seen by STB, we see the CME frontal loop as indicated by the red plus symbols. At the loop base in the lateral direction, we see a fuzzy and extended front which may correspond to the EUV wave front close to the solar limb (see also Figure 6c of \citealp{Kwon2014}), while at about 00:11~UT, only CME loops can be seen. Therefore, the EUV disturbance possibly becomes a linear wave no later than 00:16~UT. 

The SWAP telescope onboard PROBA2 also recorded this event. It has a FOV of about 1.7\,$\mathrm{R_\odot}$ and a temporal cadence of about two minutes. Therefore, it is able to capture the EUV disturbance to a larger height above the solar disk. SWAP has a passband centered at the 174\,\AA\, wavelength, peaking around $1.0\times10^6$~K. The temperature is not optimal for observing the EUV disturbance, but it can still be seen after applying the image-processing technique described in Section~3. To have the observations by STA and STB with a similar FOV to SWAP, we need to combine the EUVI and coronagraph COR 1 data. For simplicity, we denote EUVI+COR1 A and EUVI+COR1 B in the third row of Figure~\ref{fig:euv_cor_3view}. The nomenclature in other rows follows similar rules. The triple-view observations by STB, PROBA 2, and STA at about 00:20~UT are shown in the third row of Figure~\ref{fig:euv_cor_3view}. At this time, we can distinguish two fronts from the STB viewpoint. Besides the CME front indicated by the red plus signs, a second fainter front preceding the CME front can be disentangled, especially at the southern flank. Such a fainter front can be regarded as a shock wave driven by a CME as discussed by \citet{Vourlidas2003} and \citet{Ontiveros2009}. Therefore, the EUV disturbances got non-linearized and became a shock wave at about 00:20~UT. The COR 1 disturbance seems to be the upper counterpart of the EUV disturbance as revealed by the continuous transition of the shock from the EUVI to COR 1 FOV \citep{Kwon2013}. We obtain the evidence of the shock wave further from the Type II emission shown in Section~2.2 with an earliest shock signal at about 00:19:30~UT.

Further out, the CME and its driven shock wave were mainly observed by coronagraphs onboard STB, \textit{Solar and Heliospheric Observatory} (SOHO), and STA. To view CME and shock structures more completely, the coronagraph images with different FOVs are merged together as indicated in the fourth and fifth rows of Figure~\ref{fig:euv_cor_3view}. To enhance the faint shock signal, we also apply a radial filter to the coronagraph images. The CME fronts and the fainter shock fronts can be distinguished and are indicated by red and yellow arrows, respectively, in the fourth row of Figure~\ref{fig:euv_cor_3view}. Again the shock in LASCO-C2 can be traced downwards to the AIA FOV, especially close to the North Pole. In the fifth row, the CME and shock fronts are less pronounced but still visible. Note that the fronts in the last two rows of Figure~\ref{fig:euv_cor_3view} exhibit a concave shape around the central position angle of the CME and the shock.

\subsection{Radio Observations}

\begin{figure}[htbp]
\begin{center}
\includegraphics[width=1.\textwidth,clip=]{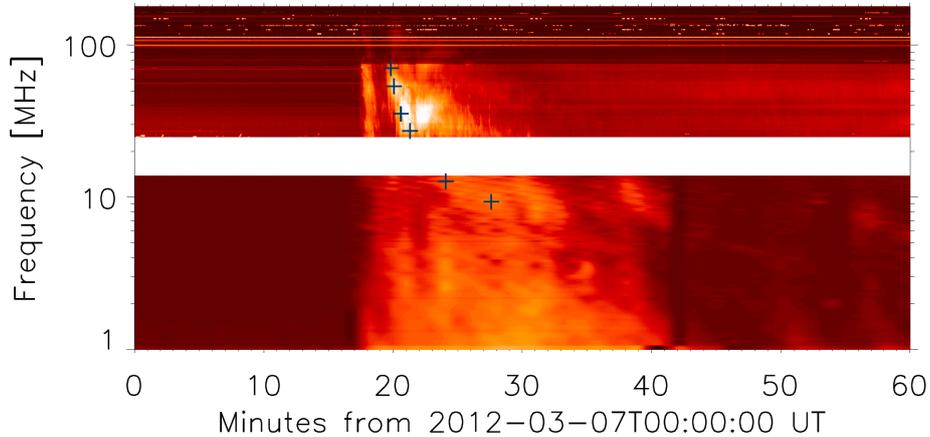}
\caption{Combined radio dynamic spectra recorded by the Learmonth station in the frequency range of 25 to 180 MHz and WAVES onboard the \textit{Wind} spacecraft in the frequency range of 1 to 16 MHz. The plus (+) symbols mark the hand-traced lane of the Type II radio burst.}
\label{fig:radio_dynamic_spectra}
\end{center}
\end{figure}

Type II radio bursts are slowly drifting structures in radio dynamic spectra \citep[e.g.][]{Vrsnak2004, Lu2017, Ying2018}. They are generally radio signals generated by shock waves. Figure~\ref{fig:radio_dynamic_spectra} shows the combined dynamic spectra measured by the Learmonth radio station in the frequency range from 25 to 180 MHz, and by \textit{Wind}/WAVES from 1 to 13.6 MHz. The plus symbols refer to the Type II lane possibly associated with the shock wave presented in Figure~\ref{fig:euv_cor_3view}. The earliest time of the Type II emission that we can recognize is about 00:19:30~UT. 

\section{Methods}

\subsection{Running Center-Median Filtering of AIA and SWAP Images}

\begin{figure}[htbp] 
\begin{center}
\hbox{
\includegraphics[width=0.48\textwidth,clip=]{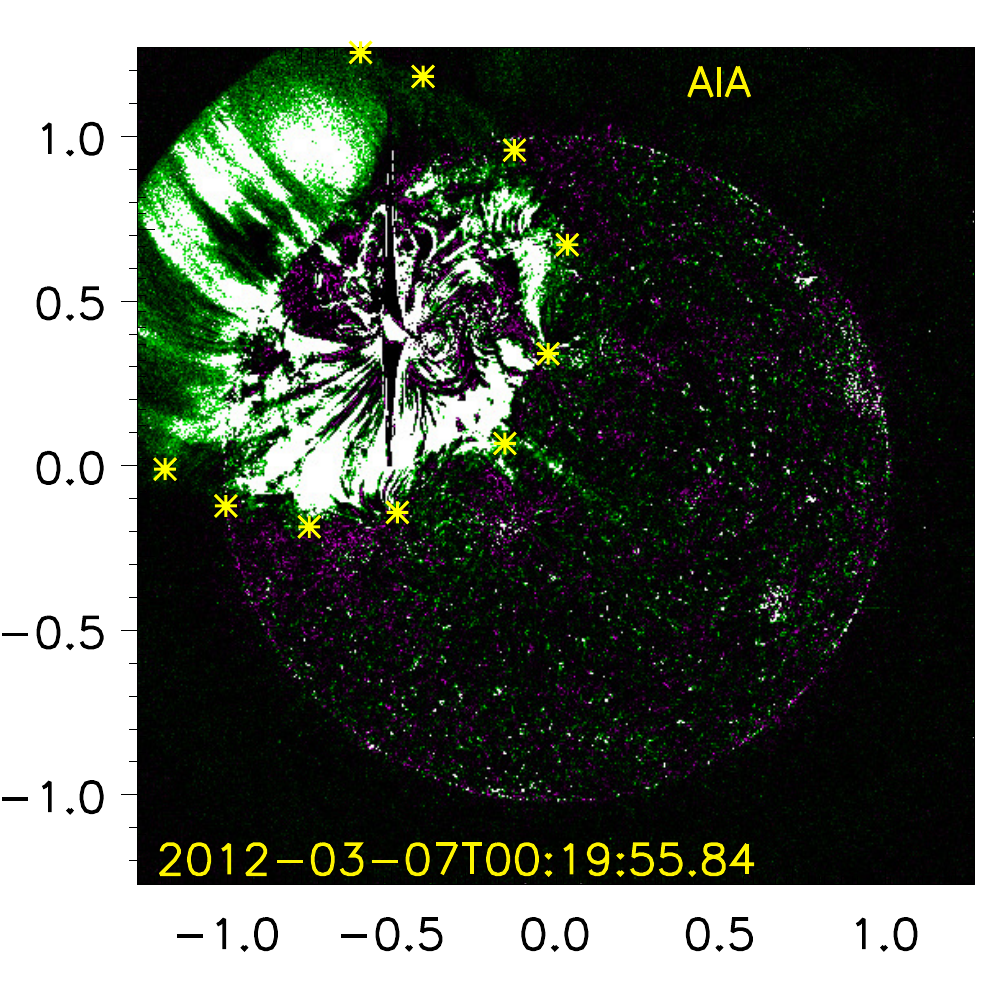}
\includegraphics[width=0.48\textwidth,clip=]{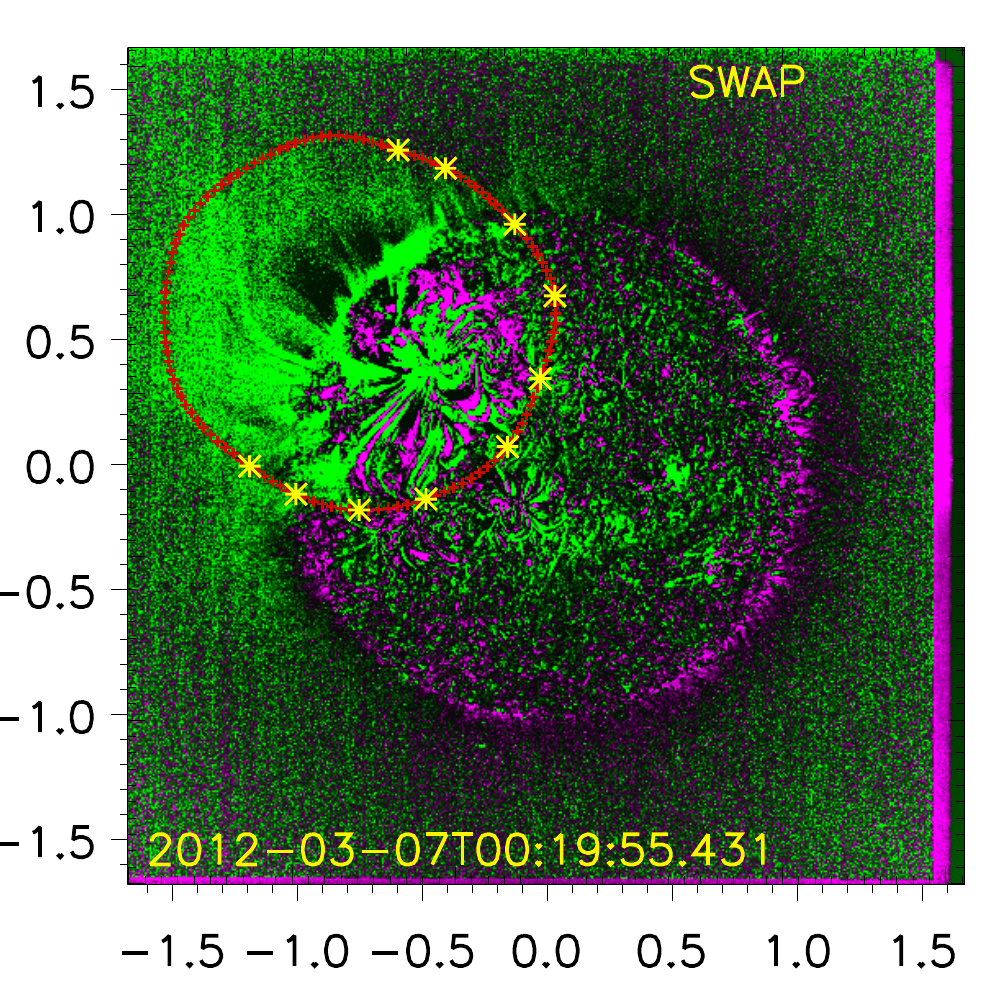}
}
\caption{Left and right: RCM filtered AIA and SWAP images at 193\,\AA\, and at 174\,\AA, respectively.  The hand-traced wave fronts from AIA and SWAP images are marked by the yellow asterisks (*) and a red curve. Green pixels indicate brightening while purple pixels indicate dimming, and a larger image intensity implies a greater change in brightening or dimming.}
\label{fig:center_median}
\end{center}
\end{figure}

We have developed a method to detect temporal variation in EUV images by applying a running center-median (RCM) filter pixel by pixel \citep{Plowman2016}. The method can identify the time-varying signals in a time sequence of images.
The RCM filter produces the per-pixel median of all images falling in a specified time window centered at a given image. Similar to the running- and base-difference images that are commonly used, our method also computes a difference which is between the given image and the produced median image.
The method works for the timescales from twice the sampling interval to about half the specified window width. For a moving feature such as our EUV disturbances, we use a sampling interval of four frames for AIA, and one frame for SWAP, while the window width was 36 frames for AIA and eight frames for SWAP. For details of the method, please refer to \citet{Plowman2016}. 

Figure~\ref{fig:center_median} shows the RCM filtered AIA and SWAP images at about 00:20:00~UT in the left and right panels, respectively. Green pixels indicate brightening while purple pixels indicate dimming, and a greater image intensity implies a more pronounced change in brightening or dimming. 
Note that due to the low temporal resolution of EUVI observations, the center-median filter is only applied to AIA and SWAP observations. We can see that the EUV disturbances are more prominent in RCM filtered images, when comparing with the corresponding structures in the base-difference images in the middle column of Figure~\ref{fig:euv_cor_3view}.
The hand-traced EUV disturbance is shown as yellow asterisks and a red curve in the AIA and SWAP images, respectively. The yellow asterisks in the SWAP image have the same coordinates of the EUV disturbance in the AIA image.

\subsection{Mask-Fitting Method for 3D Reconstruction of a Coronal Wave Surface}

\begin{figure}[htbp]
\begin{center}
\vbox{
\includegraphics[width=1\textwidth,clip=]{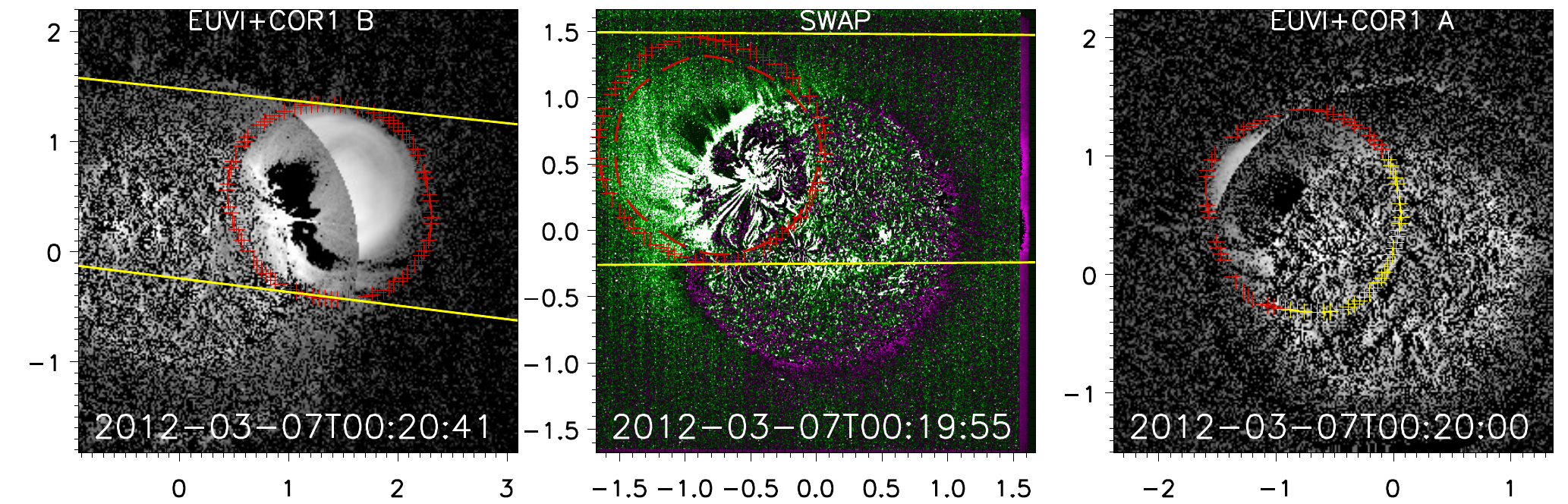}
\hbox{
\includegraphics[trim=0.2cm 1.2cm  1.5cm  1.5cm, width=0.45\textwidth,clip=]{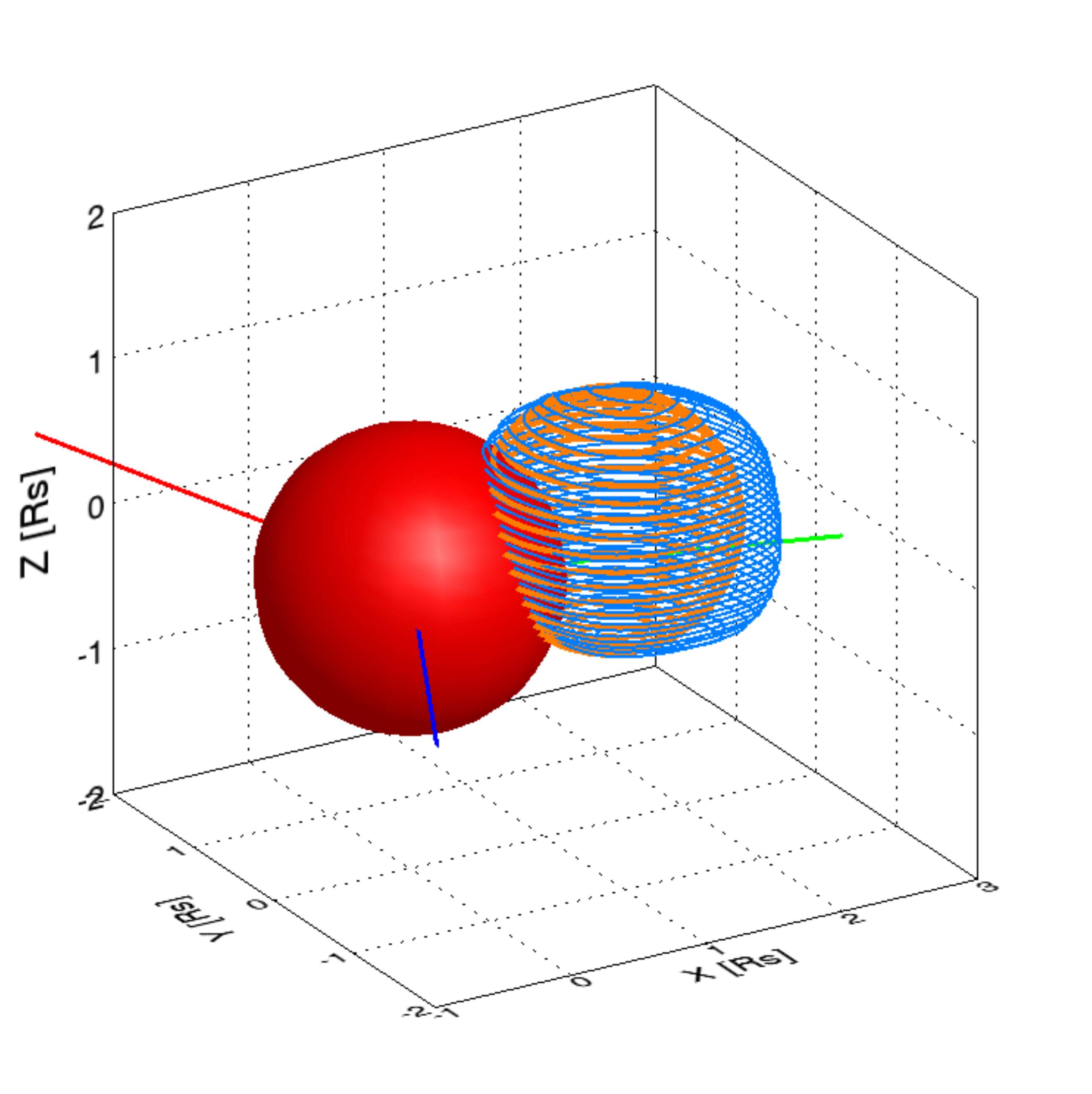}
\hspace{0.5cm}
\includegraphics[trim=2cm 3cm  4cm  3.cm, height=0.45\textwidth,clip=]{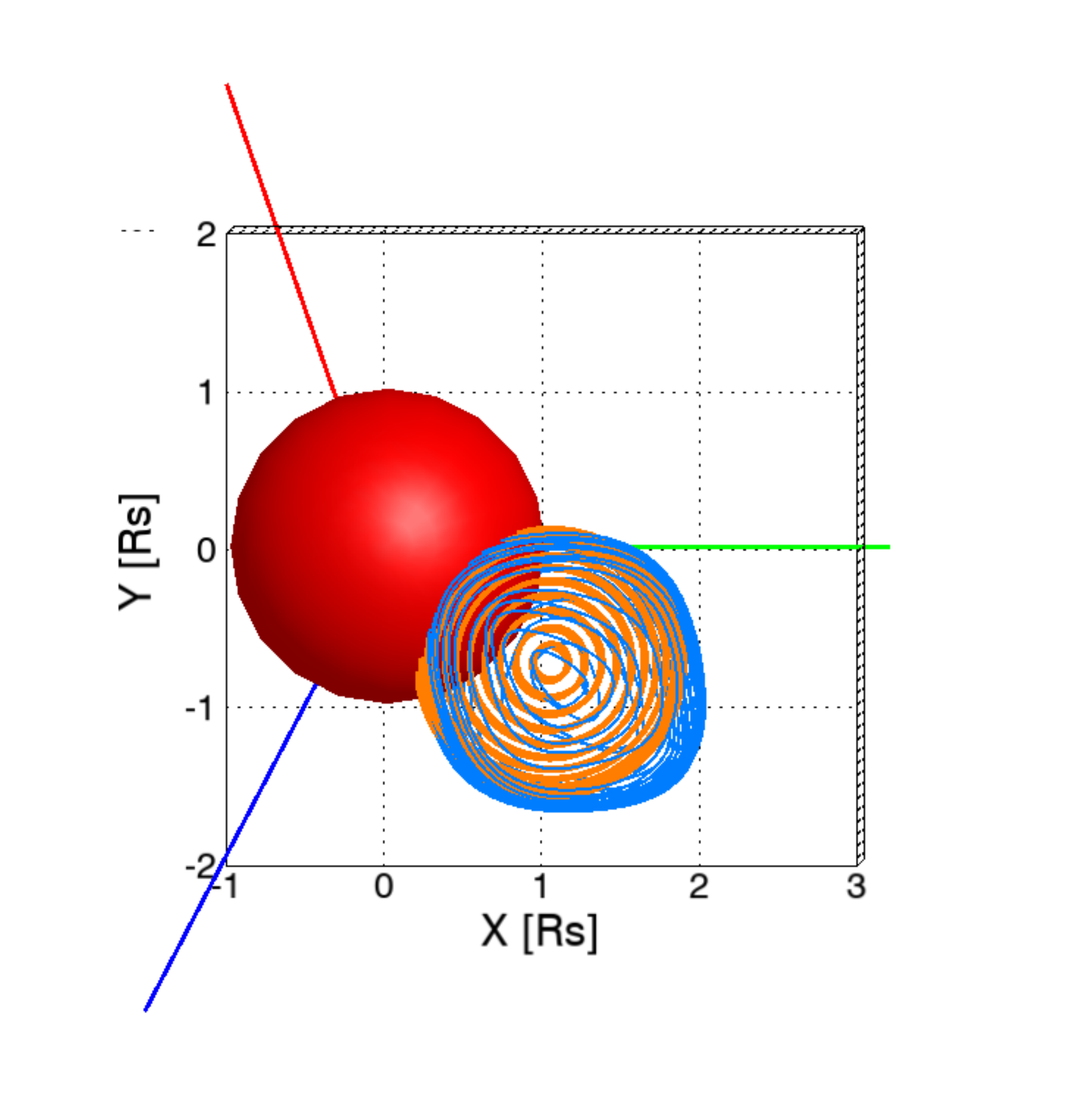}}}
\caption{Up: The traced shock periphery in the EUVI and COR1 image from STB, in SWAP, and the EUVI and COR1 image from STA. Two almost parallel yellow lines in the images observed by STB and SWAP are the uppermost and lowermost epipolar lines calculated from the shock periphery in the image observed by STA. The red-plus (+) symbols mark the hand-traced periphery within the constraint of epipolar lines. The overplotted red dashed line in the middle panel is the traced periphery in Figure~\ref{fig:center_median}. The yellow-plus (+) symbols in the right panel are the shock periphery behind the limb using the projection of the best-fit sphere as a reference. Bottom: The 3D-shocks at about 00:20:00~UT with the mask-fitting method in light-blue color and the spherical-fitting method in orange color.  The red, blue, and green straight lines indicate the view directions of STA, STB, and Earth, respectively. The red sphere represents the Sun. The left and right panels illustrate the 3D-shocks from the side and top views.}
\label{fig:mask_epi_sphere}
\end{center}
\end{figure}

In this work, we propose a new 3D-reconstruction method for wave surfaces by generalizing the application of the mask fitting method from CMEs \citep{Feng2012} to waves, and we improve the method by using epipolar geometry \citep{Inhester2006} to constrain the tracing of the wave in the images observed from different perspectives. 
The principal idea of the mask-fitting method is that we first create a mask image bounded by the periphery of the body of interest seen from an observer. The mask images obtained from multiple viewpoints demarcate a volume in a 3D-data cube.
Each discrete 3D-grid point is projected back onto the image planes. If all of the projections of a grid point are located within the masks, the grid point is labeled as a body-associated 3D-point. These points constrain the shape of the body with a hexagon on each plane parallel to the solar ecliptic plane. We smooth the hexagon using Bezier curves to finally create the 3D-volume of the body. For details of the method and its comparison with other methods, we refer readers to \citet{Feng2012}, \citet{Feng2013a}, and \citet{Feng2013b}. 
 
To improve the method, we refine the tracing of the CME/shock region by implementing the epipolar geometry as constraints. The process is demonstrated in Figure~\ref{fig:mask_epi_sphere}. We start the hand tracing of the shock periphery in the combined EUVI and COR1 image taken by STA. As its observational time is between those of EUVI+COR1 B and SWAP, we can minimize the slight shock evolution during the observational time difference from three different viewpoints. The plus symbols in the right panel of Figure~\ref{fig:mask_epi_sphere} indicate the traced shock periphery. The yellow-plus symbols are adopted from the periphery behind the limb using the projection of the best-fit sphere as a reference (see the method in Section 4.1 and the best-fit sphere in orange color in the bottom row of Figure~\ref{fig:mask_epi_sphere}). We then calculate the corresponding uppermost and lowermost epipolar lines in EUVI+COR1 B and SWAP, which are illustrated by the two almost parallel yellow lines in the left and middle panels. The tracing of the shock periphery is subsequently constrained by these two epipolar lines to compensate for the shock evolution during the observational time difference among different viewpoints. In the left and middle panels, the red-plus symbols mark the hand-traced shock wave periphery within the constraint of epipolar lines. The overplotted red dashed line in the middle panel is the traced periphery from Figure~\ref{fig:center_median}. 

The benefit of using epipolar geometry to constrain the tracing is that it compensates for the difference of about 50 seconds in observation time among the three spacecraft, and it yields a more precise 3D-volume than those derived with the version of \citet{Feng2012}. The second improvement to the mask-fitting method is that instead of using a hexagon to constrain the recorded 3D-CME points obtained from three view directions in each plane, we now adopt the convex hull (or convex envelope) method to refine such a constraint. The convex hull is the smallest convex polygon that contains a set of points. 
 The resulting 3D-shock is presented in the lower panels of Figure~\ref{fig:mask_epi_sphere} by the light-blue lines. 

\subsection{Performance of the Mask-Fitting Method}

\begin{figure}[htbp]
\begin{center}
\vbox{
\hbox{
\includegraphics[trim=0.cm 0.cm  0.cm  0.cm, width=0.47\textwidth, height=0.47\textwidth,clip=]{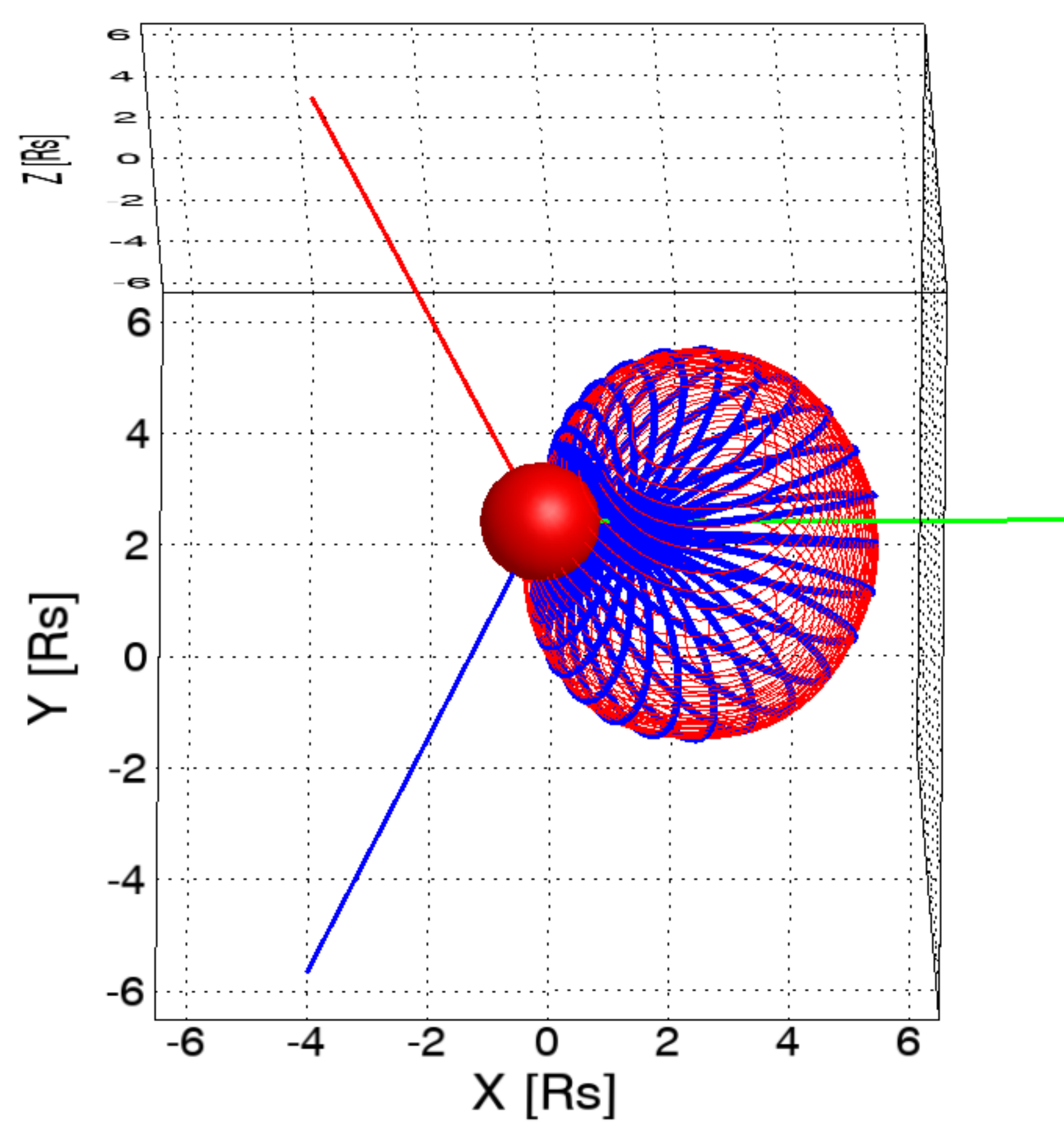}
\hspace{0.2cm}
\includegraphics[trim=0.5cm 0.cm  0.5cm  2.0cm, width=0.48\textwidth, height=0.47\textwidth,clip=]{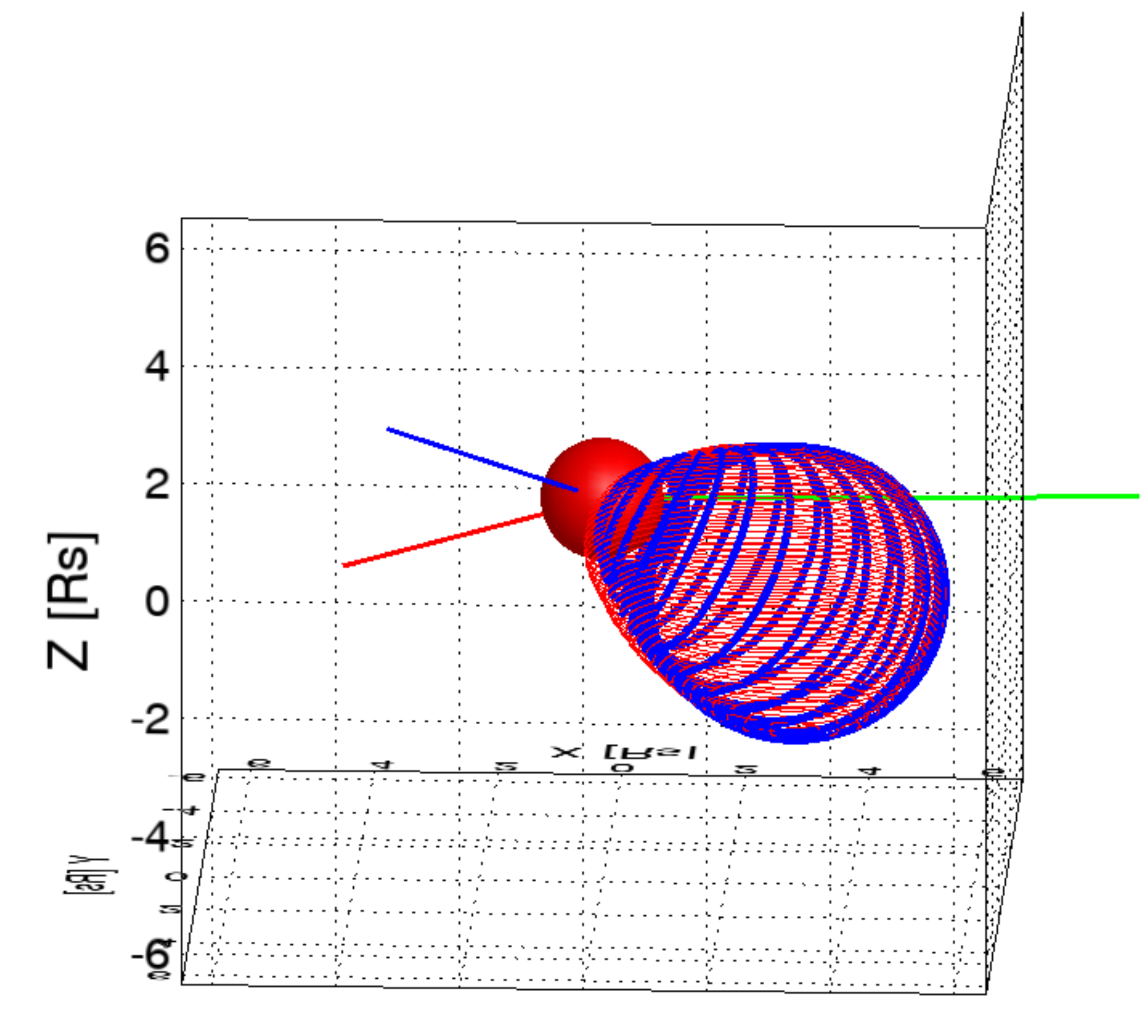}}
\hspace{0.2cm}
\includegraphics[width=1\textwidth,clip=]{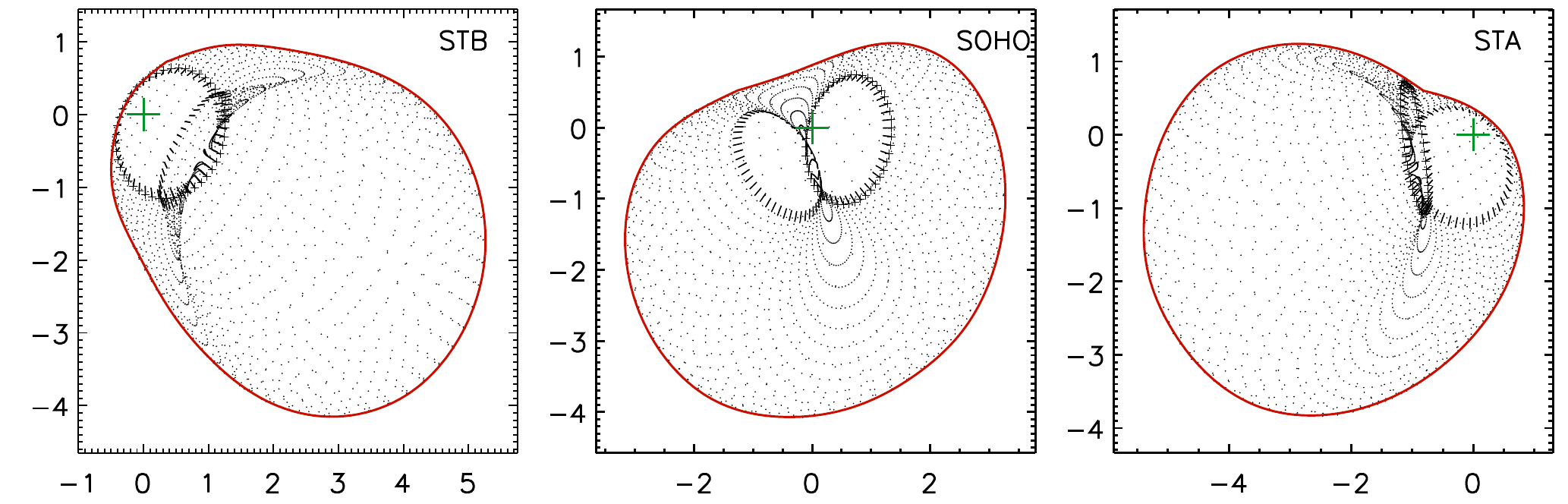}}
\caption{Top: Comparison of the simulated 3D-GCS model in blue color and the reconstructed 3D-shape in red color using the mask-fitting method. The red, blue, green straight lines, and the red sphere have the same meaning as in Figure~\ref{fig:mask_epi_sphere}. The left and right panels illustrate the comparisons from the side and top views, respectively. Bottom: projections of the 3D-GCS model onto the images observed from three viewpoints of STB, SOHO, and STA with a separation angle of about 120 degrees from each other. Black-dotted lines represent the projections, and red curves are the traced peripheries. Green-plus symbols indicate the position the Sun. All of the axes are in units of the solar radius.}
\label{fig:validate_mask}
\end{center}
\end{figure}

To evaluate the performance of our mask fitting method, we use a 3D-gradual cylindrical shell (GCS) model (blue color in the upper panels of Figure~\ref{fig:validate_mask}) to test whether the mask fitting method is able to reproduce a given shape.
The reason why we choose the GCS model is that it is relatively complicated and has two principal curvatures around its front, which can differ significantly. From the side and top views, the larger and smaller principal curvatures of the GCS front can be qualitatively compared. In the lower panels of Figure~\ref{fig:validate_mask}, we project the simulated 3D-GCS model onto the image planes that are separated by about 120 degrees from one another. The projections are indicated by the black-dotted lines. The tracing of peripheries uses the epipolar lines as described in Section 3.2 and is delineated by red curves. The reconstructed 3D-surface using the mask-fitting method is displayed by the red curves in the upper panels.  As revealed by the comparison between the model and the reconstructed 3D-surface, the method can capture the varying curvatures along the GCS structure well within an average distance error of 0.09 $\mathrm{R_\odot}$ between the model and the reconstructed surface.

\section{Results and Discussions} 

\subsection{Model-free 3D Surface of Coronal Wave Fronts}

\begin{figure}[htbp]
\begin{center}
\vbox{
\includegraphics[trim=1.0cm 1.0cm  1.0cm  1.0cm, width=1\textwidth,clip=]{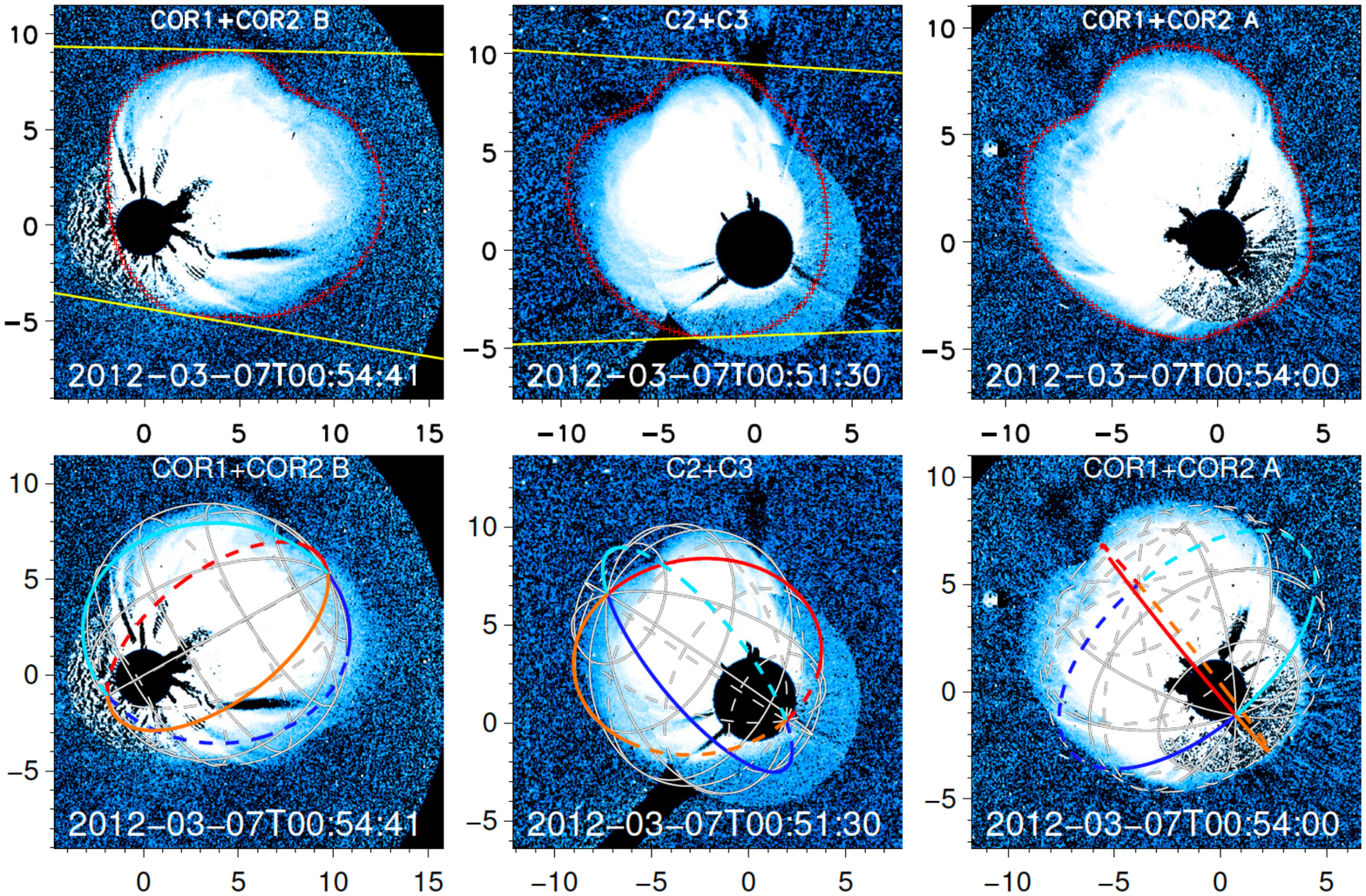}
\hbox{
\includegraphics[trim=0.cm 1.0cm  1.5cm  1.0cm, width=0.47\textwidth,clip=]{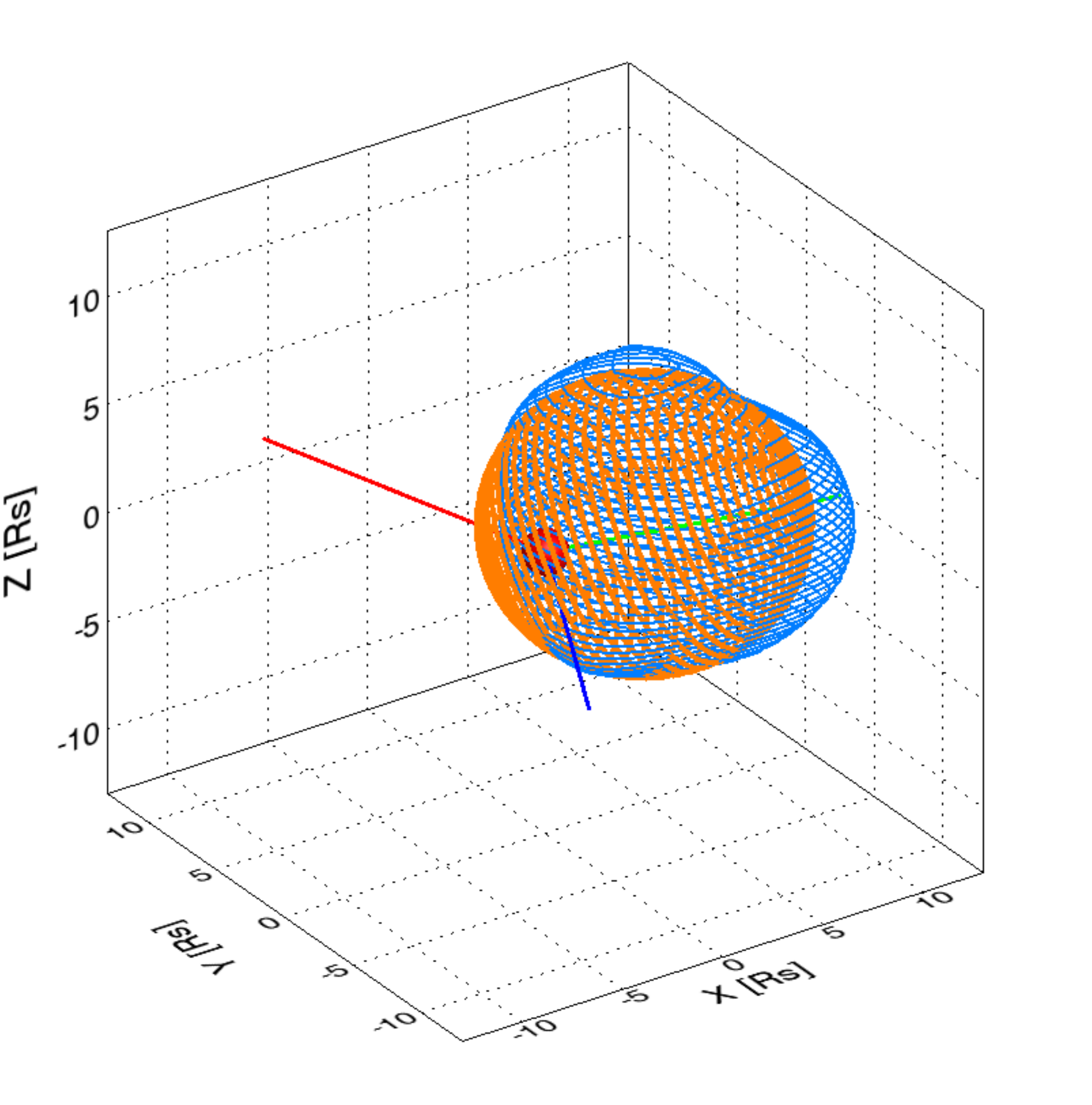}
\hspace{0.2cm}
\includegraphics[trim=0.cm 0.cm  0.cm  0.cm, height=0.47\textwidth,clip=]{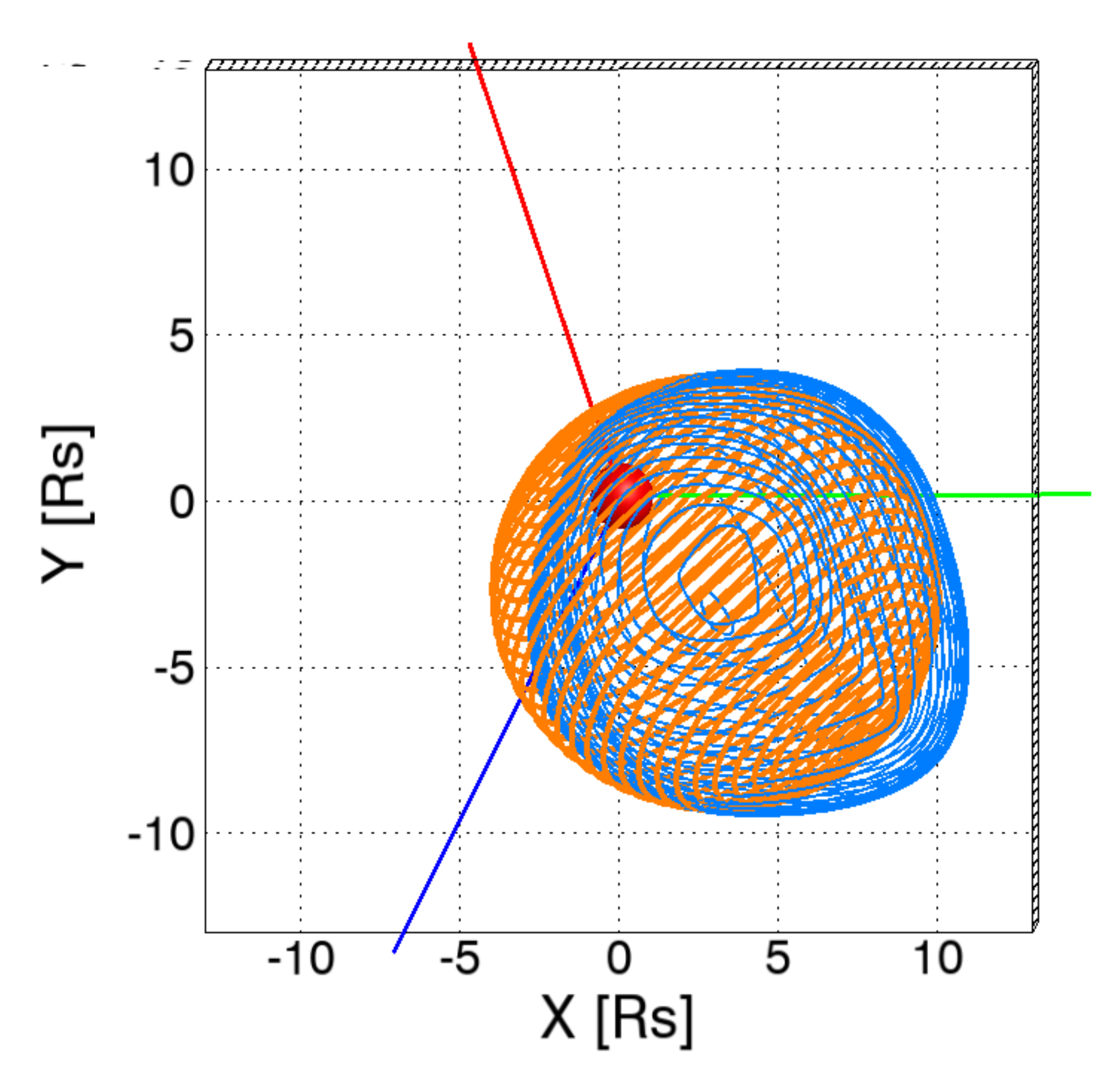}}
}
\caption{Top: The hand-traced shock peripheries are delineated by red lines in the images from three perspectives. The calculated epipolar ranges in COR1+COR2 B and C2+C3 images from the tracing in COR1+COR2 A are indicated by yellow lines. The projections of the reconstructed 3D-shock surface onto three images using the mask-fitting method are illustrated by solid lines in orange color.  Middle: projections of the best-fit ellipsoid model onto the coronagraph observations from STB, SOHO, and STA at about 00:54~UT. All of the axes are in units of the solar radius. Bottom:  Comparison of the 3D-shock at about 00:54:00~UT with the ellipsoidal-fitting method in orange, and the mask-fitting method in light blue. The left and right panels illustrate the comparisons from the side and top views, respectively.}
\label{fig:proj_3d_0054}
\end{center}
\end{figure}

The mask fitting method offers a way to derive a model-free 3D-surface of an observed body, while the forward-fitting methods commonly used to determine the 3D-geometry of CME and shocks \citep[e.g.][]{Kouloumvakos2016,Kwon2014} assume idealized shapes of the structures, e.g. GCS, sphere, or ellipsoid.

We find epipolar lines corresponding to the shock periphery on the COR1+COR2 A image at about 00:54~UT. The two yellow lines in the COR1+COR2 B and C2+C3 images in the first row of Figure~\ref{fig:proj_3d_0054} are the epipolar lines when they are tangent to the peripheries in these images.  In this way, epipolar lines calculated from an image constrain in the 3D-space the peripheries of the shock projected onto the other images. Since the method assumes that the images are exactly co-temporal, we compensate the time difference by assuming a self-similar expansion of the peripheries in the image planes. The observation time of the shock in the C2+C3 image is about 2.5 minutes earlier than the time in COR1+COR2 A. We use the traced shock periphery in the next frame and shrink it self-similarly to fill the space between the two epipolar lines. Then the shrunken shock periphery is utilized as a reference to trace the shock in C2+C3. 
The tracing results are represented by red symbols. The side and top views of the resulting 3D-shock surface are shown in light-blue color in the third row of Figure~\ref{fig:proj_3d_0054}. The concave structure along the shock front is successfully captured from the side view. We also project the 3D-shock surface onto the images from three perspectives. The projections are presented by the solid lines in orange color in the first row of Figure~\ref{fig:proj_3d_0054}. We can see that the projections fit the trace peripheries almost perfectly.

In the second row of Figure~\ref{fig:proj_3d_0054}, we show the projections of the best-fit ellipsoid onto the three images. They can overlap the observed shock in general. Unfortunately, they cannot fit the concave structures of the shock shape.
In the third row of Figure~\ref{fig:proj_3d_0054}, we also plot the 3D-shock ellipsoid in orange color. From the comparison of the results derived from the ellipsoidal-fitting and mask-fitting methods, we reveal that both methods produce a consistent shock direction of about 35 degrees east of the direction to the Earth, especially from the top view. Unlike the ellipsoid model, the mask fitting method derives the model-free 3D-surface of the shock, i.e., the concave shock front from the side view, and a larger curvature around the shock nose from the top view. Therefore the mask-fitting method can reflect the highly inhomogeneous coronal medium in which the shock propagates. For associating shock key parameters (e.g. shock normal to magnetic field direction) to SEP properties, such a more realistic curvature is required.

\subsection{Morphological Evolution}

\begin{figure}[htbp]
\begin{center}
\hbox{
\includegraphics[trim=0.5cm 0.5cm 0.8cm 0.5cm, width=0.49\textwidth,clip=]{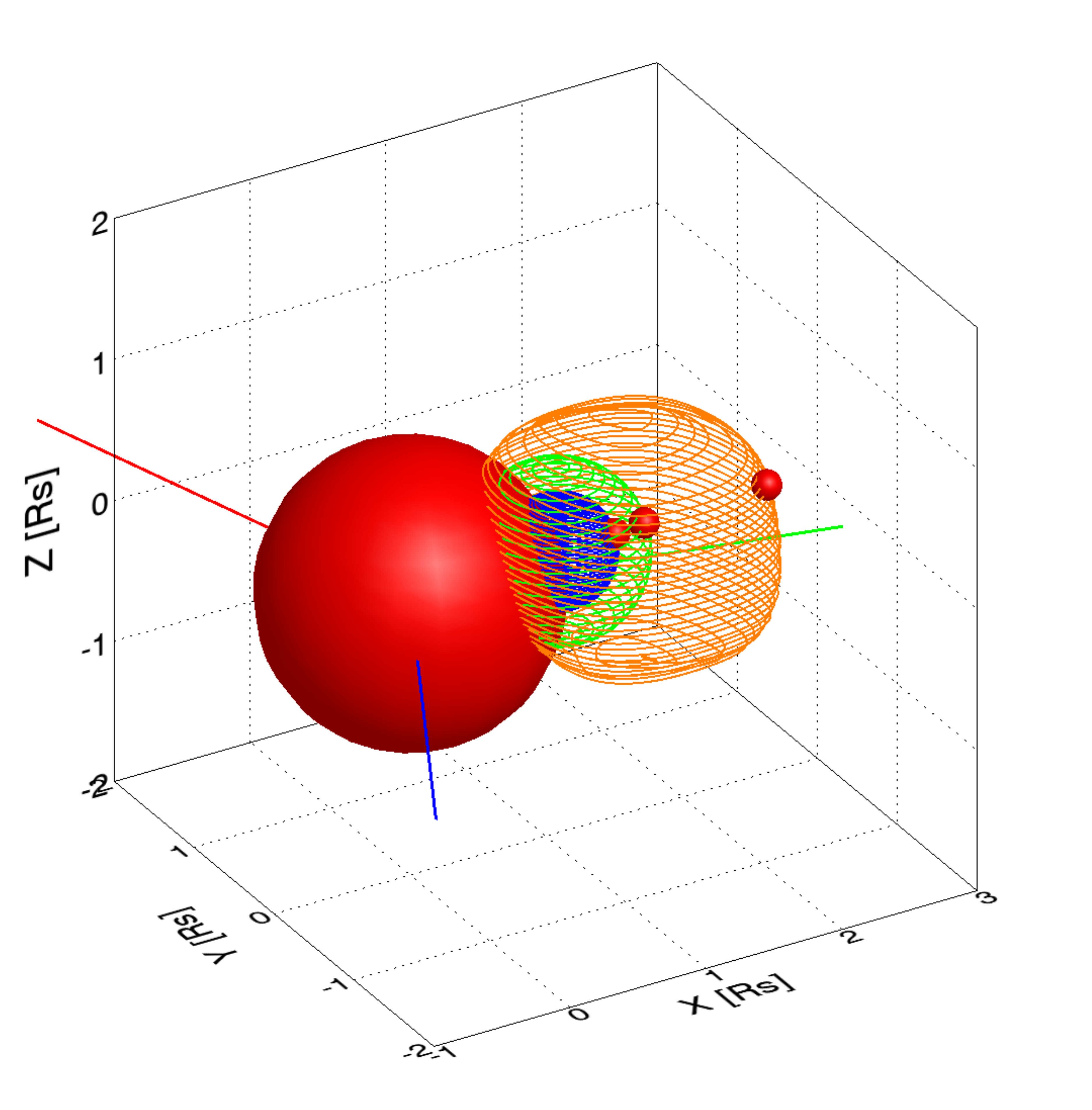}
\includegraphics[trim=0.cm 0.5cm 1.0cm 0.5cm, width=0.48\textwidth,clip=]{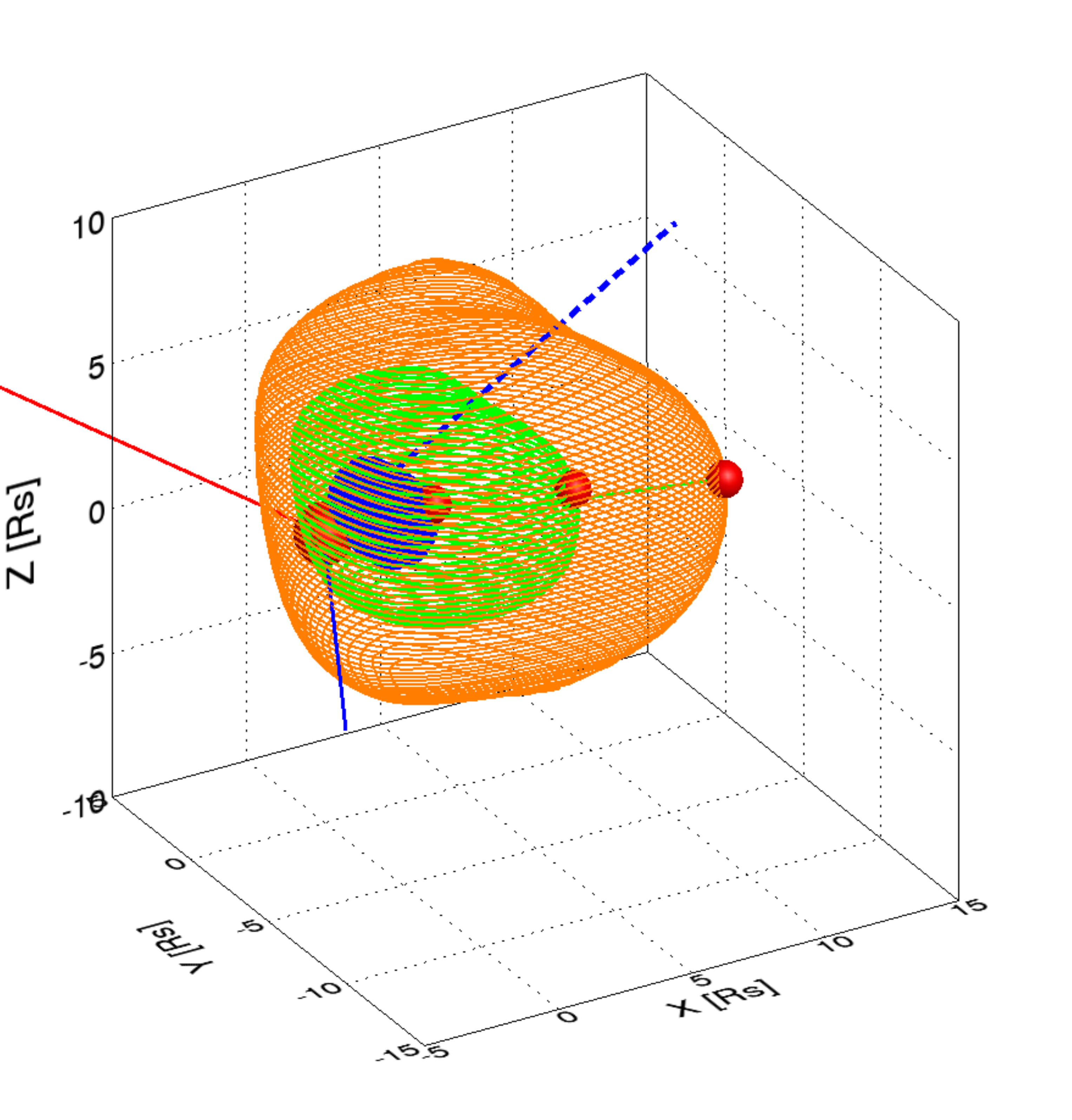}}
\caption{Left: 3D-reconstructions using spherical fitting at about 00:11~UT in blue color, at about 00:16~UT in green color, using mask fitting at about 00:20~UT in orange color. Right: 3D-reconstructions using mask fitting at about 00:25~UT, 00:39~UT, and 00:54~UT in blue, green, and orange colors, respectively. The red, blue, green straight lines, and the bigger red sphere have the same meaning as in Figure~\ref{fig:mask_epi_sphere}, and the smaller red spheres mark the nose position. The blue dashed line indicate the direction in which the concave structure gradually develops.}
\label{fig:morphology}
\end{center}
\end{figure}

Figure~\ref{fig:morphology} shows the reconstructed 3D-surfaces from about 00:11~UT to 00:20~UT in the left panel, and from about 00:20~UT to 00:54~UT in the right panel. For the first two time frames of the reconstructions from EUV observations, we adopt the spherical-fitting method, as the mask-fitting method is designed for larger FOV coronagraph observations. Following the morphological evolution of the 3D-surface, we can see the development of the concave structure in the direction indicated by the blue-dashed line in the right panel of Figure~\ref{fig:morphology}. It is barely visible until 00:25~UT. At 00:39~UT, the front along this direction become flattened. At 00:54~UT, the concave structure is clearly seen. 
We have checked the original coronagraph images without applying base differences, and found coronal streamers locate around the concave structure in the original 2D-images. One of the interpretations of the concave structure in 3D may be due to the interaction of the wave with the streamer belt. Due to the high density, the fast magnetosonic speed in the streamer would be lower than its surroundings. In Figure~\ref{fig:morphology} we also marked the nose position of the 3D-surfaces by the small red spheres. Note that the surface at 00:11~UT may be the CME frontal loop rather than a wave surface, as already discussed in Section~2.1. This idea is further supported by the very low speed analyzed in Section~4.3 below.

\subsection{Kinematic Evolution}

\begin{figure}[htbp]
\begin{center}
\includegraphics[width=0.8\textwidth,clip=]{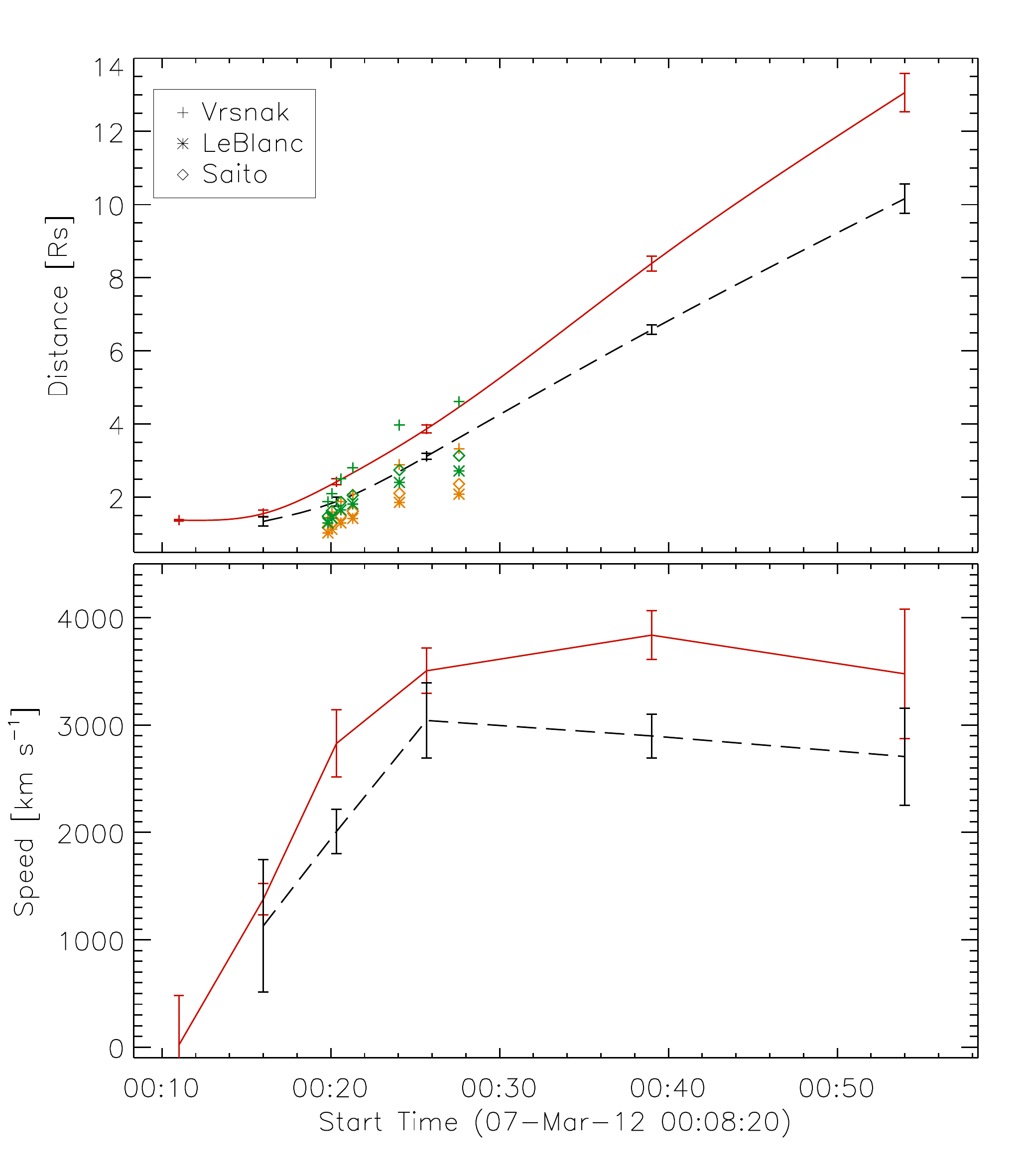}
\caption{Top: Distance of the wave nose and the concave structure as a function of time derived from the 3D-wave surfaces are indicated by the red and black symbols. The red-solid and black-dashed curves are their corresponding spline interpolations. The associated error bars are propagated from the uncertainty of the wave tracing in images observed from three perspectives. The plus (+), asterisk (*), and diamond symbols are inverted distances from Type II frequencies using density models by
\citet{Saito1977}, \citet{Leblanc1998}, and \citet{Vrsnak2004}.
The orange and green symbols correspond to the inverted distance by assuming the Type II burst are at fundamental and second-harmonic frequencies, respectively. Bottom: Speed of the wave nose and the concave structure as a function of time are indicated in red and black colors, respectively. The error bars are propagated from the uncertainty of distance.}
\label{fig:wave_ht_nose_typeii}
\end{center}
\end{figure}

In this subsection, we determine the wave kinematics in different wavebands, at different heights, and in different directions. The kinematics derived from the AIA 193\,\AA\, RCM images and from the 3D-reconstructions offer the kinematics on the solar disk and in the extended corona beyond the limb, respectively. The Type II radio burst also could provide us with the shock kinematics, although it is difficult to locate the radio source in the reconstructed 3D-shock surface without radio imaging observations. 

In the upper panel of Figure~\ref{fig:wave_ht_nose_typeii}, the red symbols refer to the distance of the wave nose, which is the furthest point from the solar center. The error bars are propagated from the uncertainty of the wave tracing. For a given time, we repeat the tracing of wave peripheries in three images independently for eight times, then we apply the mask-fitting methods for each set of three tracings. The uncertainty of the nose distance are the 3$\sigma$ of the calculated distances at the eight times. We also smooth the averaged distance over the computations in eight times using splines as indicated by the red curve in the upper panel of Figure~\ref{fig:wave_ht_nose_typeii}. The resulting speed is presented in the lower panel of Figure~\ref{fig:wave_ht_nose_typeii} with error bars shown in red, which are propagated from the uncertainty of distances using Monte-Carlo simulations. At each distance--time data point, we randomly select 100 distance values which follow a Gaussian distribution with $3\sigma$ equals to the 3$\sigma$ of the calculated distances.
The speeds are correspondingly computed 100 times, and the mean and $3\sigma$ are plotted in the lower panel. The speed increases from a very low value below a few hundred km~s$^{-1}$ at 00:11~UT to a maximum value of about 3800 km~s$^{-1}$ at 00:39~UT, and it starts to decrease afterwards. The maximum speed is about 25\,\% higher than that found by \citet{Kwon2014}. As a comparison, the kinematics in the direction of the concave structure is included as well in Figure~\ref{fig:wave_ht_nose_typeii} in black color. 
Note that the data point at 00:11~UT is not included, because in Figure~\ref{fig:morphology} the blue dashed line does not intersect with the wave surface at this time.
It is obvious that the wave starts to decelerate earlier, and has a significantly lower speed in this direction compared with the case in the nose direction.

According to the plasma emission mechanism of the Type II burst, the frequency corresponds to the local electron density. The electron density $n_\mathrm{e}$ in units of cm$^{-3}$ are obtained by using $f_\mathrm{p}=9 \times10^{-3}\sqrt{n_\mathrm{e}}$, where $f_\mathrm{p}$ is the fundamental plasma frequency in units of MHz. 
Since it is not certain if the emissions are at the fundamental frequencies, both fundamental and second-harmonic frequencies are assumed when calculating the electron densities.
To obtain a distance--time plot, the derived electron density is often converted to radial distance by using an analytical background-density model. 
We adopt three different electron density models given by \citet{Saito1977}, \citet{Leblanc1998}, and \citet{Vrsnak2004} to estimate the distance of the shock from the solar center. In the upper panel of Figure~\ref{fig:wave_ht_nose_typeii}, the computed distances of Type II sources are plotted as orange and green symbols by assuming the emissions are at fundamental- and second-harmonic frequencies, respectively. The plus, asterisk, and diamond symbols represent the distances derived with different density models. By comparing the distance of the shock nose and Type II source, we find that it is probable that at the beginning of the Type II burst at about 00:20~UT, its source region is located at the shock flank.  However, to confirm this, radio-imaging observations \citep[e.g.][]{ChenY2014, Morosan2019} are required. At later time, due to large scattering of the distances in green and orange, we cannot draw any conclusion about the source location. The large scattering of the distances also yields a wide range of average speed from about 1600\,km~s$^{-1}$ to close to 4000\,km~s$^{-1}$ using a linear fit to each distance--time plot.

\begin{figure}[htbp]
\begin{center}
\vbox
{\includegraphics[width=0.55\textwidth,clip=]{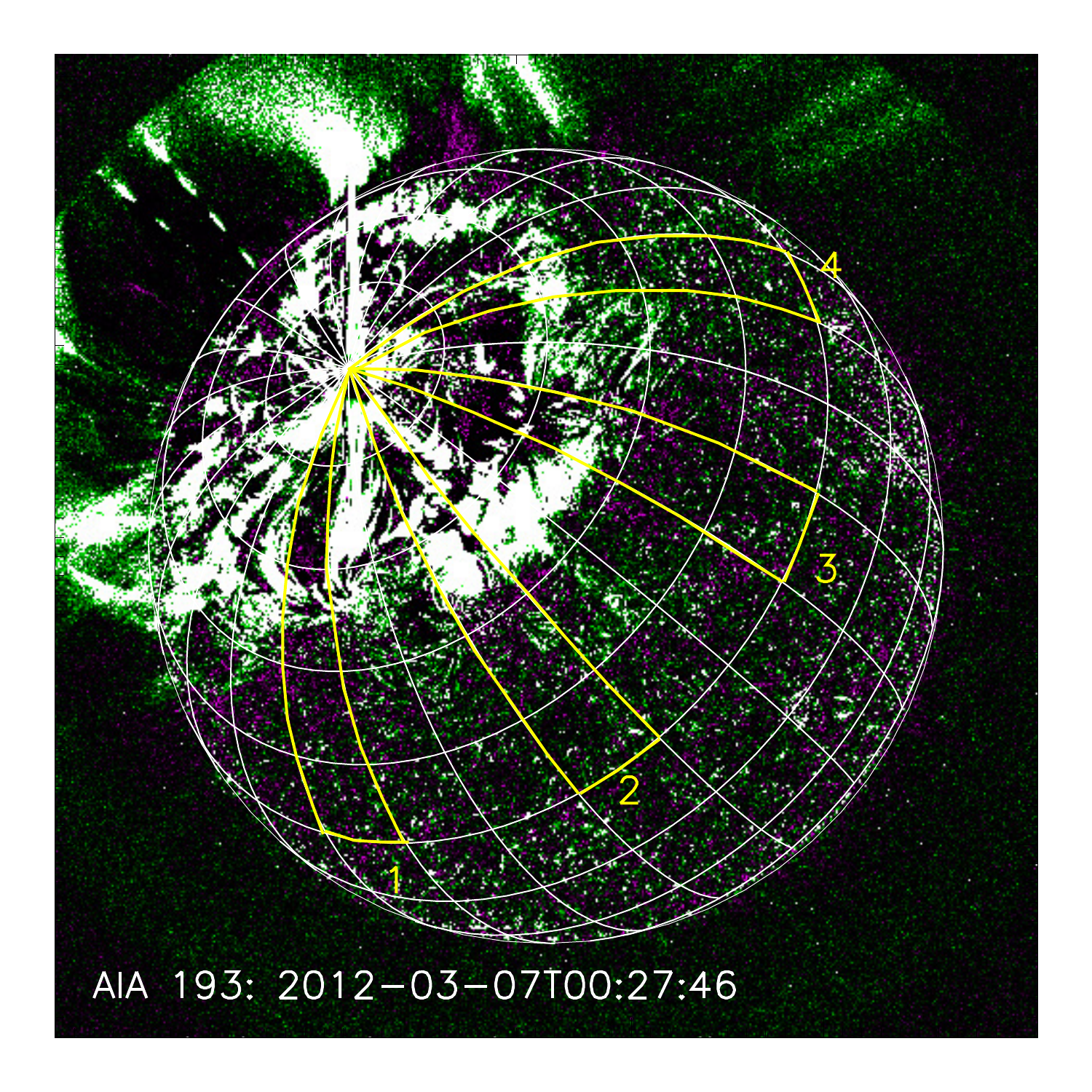}
\includegraphics[width=1\textwidth,clip=]{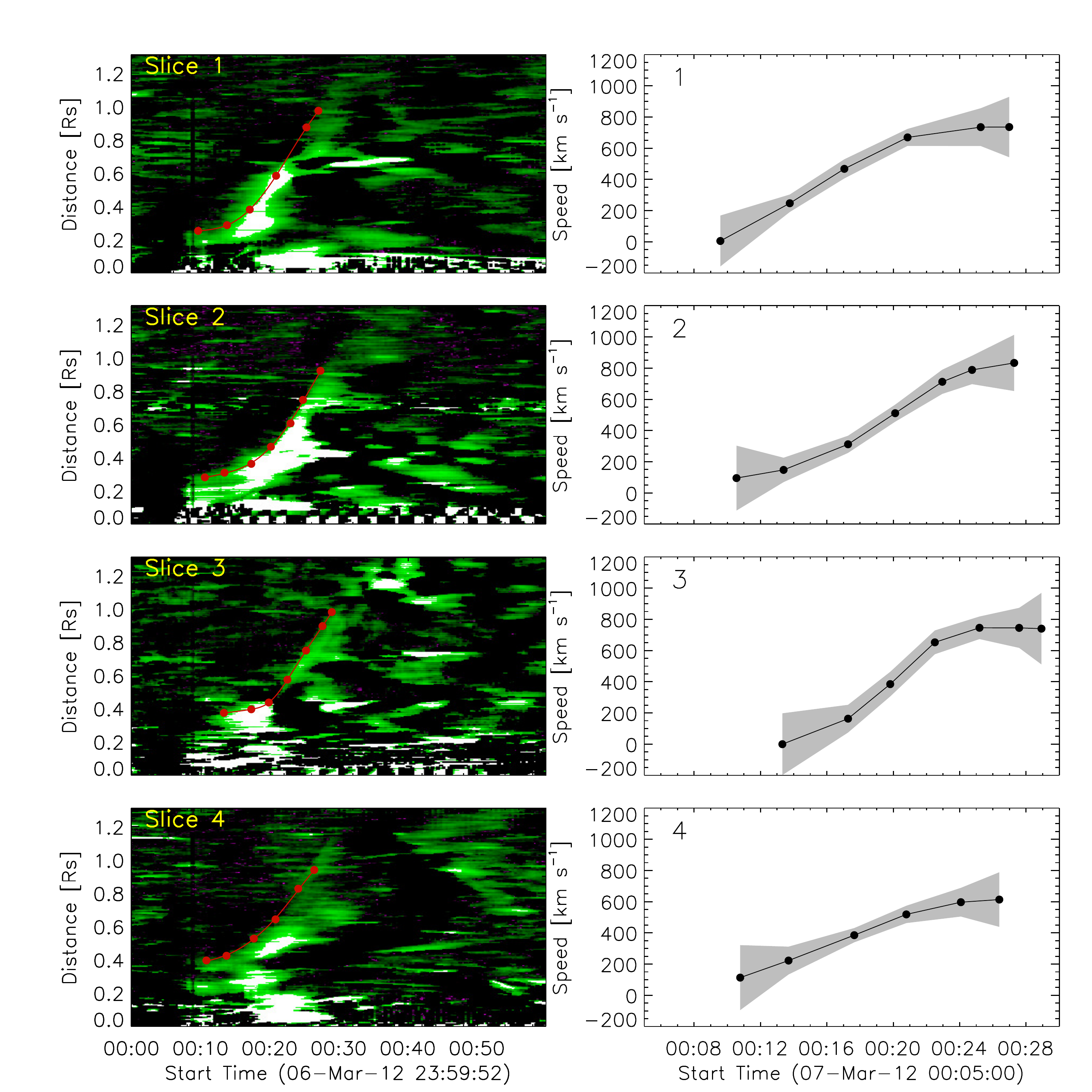}}
\caption{Upper: selected sectors 1\,--\,4 that are used to derive the distance--time plots of the EUV disturbances are indicated by yellow lines. The longitudinal and latitudinal grids with the pole centered at the eruption center are indicated by white lines. Bottom: distance--time plots (left) and speed--time plots (right) for sectors 1\,--\,4 from top to bottom panels. The error bars are calculated under the assumption that the uncertainty in wave front positions is 0.05 $\mathrm{R_\odot}$. }
\label{fig:wave_kinem_aia}
\end{center}
\end{figure}

To compare the speed of wave nose to the speed of the global EUV disturbances across the solar disk, we have analyzed AIA images. To derive its kinematics, we need to correct for the curvature of the solar surface, and we adopt the technique described by \citet{Nitta2013}.  First the eruption center is regarded as a pole. Then from the pole, the whole solar surface is divided into 24 equally spaced sectors in longitude. The selected four sectors and the longitudinal and latitudinal grids are presented in Figure~\ref{fig:wave_kinem_aia}. For each of the four sectors, at a given time the intensity as a function of the distance along the longitude is calculated by averaging the intensity of the pixels along the latitude. The corresponding distance--time plots for these four sectors are also shown in Figure~\ref{fig:wave_kinem_aia}. Along all four sectors, the speed of the EUV disturbances starts with a very low value and increases to about 600 to 800\,km~s$^{-1}$ from about 00:10~UT to about 00:28~UT. The shaded regions are errors obtained from the Monte-Carlo simulations with the uncertainty in the wave front positions of 0.05 $\mathrm{R_\odot}$ which is the maximum position error we can read in Figure~\ref{fig:wave_kinem_aia}.
Because of the very low starting speed, it is possible that the EUV disturbances at the beginning are not a wave but due to the magnetic reconfiguration. In this article, we infer that the speed of the EUV disturbances across the solar disk is much slower than the wave speed in the extended corona (see similar results by \citet{Kwon2017} for three other different events) and also the speed of the Type II sources.

\section{Conclusions and Outlook} 
We have investigated the coronal wave on 7 March 2012 observed from three perspectives from STA, STB, and Earth/SOHO. Taking advantage of the large FOV of SWAP, which extends to about 1.7~$\mathrm{R_\odot}$, we have a more complete view of the EUV disturbances. The RCM technique was applied to enhance the signal of EUV disturbances in AIA and SWAP images. 
We have made 3D-reconstructions of the coronal wave surface using three techniques including spherical fitting, ellipsoidal fitting, and mask fitting. The last one is a new method to derive the 3D-wave surface, which is adopted and improved from the technique originally designed for a 3D-CME surface. All of the three methods produce consistent propagation directions of the coronal wave. However, both forward-fitting methods fail to reconstruct the detailed structure of the wave surface, for example, the concave structure along the shock front, which are captured by the mask-fitting method. 

Based on the obtained 3D-wave surfaces, we have studied its morphological evolution and followed the developing of the concave structure around the shock front. It is barely visible until 00:25~UT. At 00:39~UT, the front becomes flattened. At 00:54~UT, the concave structure is clearly seen. One interpretation of the concave structure may be that it is due to the interaction of the wave with the streamer belt. The kinematics of the wave nose and the concave structure are also computed.
The nose speed increases from a low value below a few hundred km~s$^{-1}$ at 00:11~UT to a maximum value of about 3800\,km~s$^{-1}$ at 00:11~UT, and it starts to decrease afterwards. The 3D-wave in the direction of the concave structure has significantly lower speeds and starts to decelerate earlier. The kinematics of the EUV disturbances observed by AIA shows a very slow initial speed below 100\,km~s$^{-1}$, then it increases to about 600 to 800\,km~s$^{-1}$ along different paths from about 00:10~UT to 00:28~UT. In viewing of the very low initial speed, we speculate that the propagating EUV disturbances may be due to magnetic reconfiguration in the beginning. We also find that the wave in the extended corona has a much higher speed than the speed of EUV disturbances across the solar disk. 

A precise 3D-shock surface and thus derived shock parameters, e.g. kinematics, Mach number, and the angle between shock normal and upstream magnetic field, is crucial for understanding their link to the SEP properties. Furthermore, the improved mask-fitting method could be applied to the multi-perspective observations by different coronagraphs (e.g. SECCHI-COR1 and -COR2, and LASCO-C2 and -C3) and heliospheric imagers (e.g. SECCHI-HI1,  \textit{Wide-Field Imager for Parker Solar Probe} (WISPR), and \textit{Solar Orbiter Heliospheric Imager} (SoloHI). In the future, such attempts will be made towards this direction to follow the evolution of the 3D-parameters in a much larger FOV and their link to SEPs.

\acknowledgments
We are grateful to the reviewer for the constructive comments which helps us to improve the manuscript. We thank Ryun-Young Kwon for providing us with his code of the ellipsoid model.
We acknowledge the PROBA 2 Guest Investigator Program.
SWAP is a project of the Centre Spatial de Li\`{e}ge and the Royal Observatory of Belgium funded by the Belgian Federal Science Policy Office (BELSPO).
Data courtesy of NASA/SDO and the AIA science team.  SDO is a mission of NASA's Living With a Star Program.
SOHO is a project of international cooperation between NASA and ESA.
The STEREO/SECCHI data used here were produced by an international 
consortium of the Naval Research Laboratory (\,USA\,), Lockheed Martin 
Solar and Astrophysics Lab (\,USA\,), NASA Goddard Space Flight Center 
(\,USA\,), Rutherford Appleton Laboratory (\,UK\,), University of 
Birmingham (\,UK\,), Max-Planck-Institut for Solar System Research 
(\,Germany\,), Centre Spatiale de Li\`ege (\,Belgium\,), Institut d'Optique 
Th\'eorique et Appliqu\'ee (\,France\,), and Institut d'Astrophysique Spatiale 
(\,France\,). The USA institutions were funded by NASA, the UK institutions 
by the Science \& Technology Facility Council (\,which used to be the 
Particle Physics and Astronomy Research Council, PPARC\,), the German 
institutions by Deutsches Zentrum f\"{u}r Luft- und Raumfahrt e.V. (\,DLR\,), 
the Belgian institutions by Belgian Science Policy Office, and the French 
institutions by Centre National d'Etudes Spatiales (\,CNES\,) and the 
Centre National de la Recherche Scientifique (\,CNRS\,). The NRL effort was 
also supported by the USAF Space Test Program and the Office of Naval 
Research.
This work is supported by National Key Research and Development program 2018YFA0404202,
the mobility program (M-0068) of the Sino--German Science Center,
NSFC grants U1731241, 11921003, and 11973012,
and by the CAS Strategic Pioneer Program on Space Science, Grant No. XDA15052200, XDA15320103, and XDA15320301.
The development of the 3-D ellipsoid models was supported by the National Research Foundation of Korea (NRF-2019R1F1A1062079) grant funded by the Korea government (MSIT; Project No. 2019-2-850-09).

\section*{Disclosure of Potential Conflicts of Interest}
The authors declare that they have no conflicts of interest.
     
\bibliographystyle{spr-mp-sola}
\bibliography{ms.bib}

\end{article} 

\end{document}